\documentclass[%
superscriptaddress,
twocolumn,
nofootinbib,
amsmath,
amssymb,
aps,
prx,
10pt
]{revtex4-2}

\usepackage[utf8]{inputenc}
\usepackage[T1]{fontenc}
\usepackage{lmodern}                 %
\usepackage{graphicx}
\usepackage[caption=false,font=footnotesize]{subfig}
\usepackage{longtable,booktabs}
\usepackage{bbm}

\usepackage[colorlinks=true,linkcolor=teal,citecolor=Maroon,urlcolor=Maroon]{hyperref}

\newcommand{\wt}{\mathrm{wt}}

\usepackage{booktabs}    %
\usepackage{multirow}    %
\usepackage{array}       %

\newcommand{\condref}[1]{\hyperref[#1]{Condition~\ref*{#1}}}
\usepackage[x11names, svgnames, rgb, table, dvipsnames]{xcolor}
\usepackage{tom}
\usetikzlibrary{calc}
\usetikzlibrary{arrows.meta,trees, quotes, matrix, positioning, shapes, fit, calc}
\useyquantlanguage{groups}
\usepackage[caption=false,font=footnotesize]{subfig}  %
\makeatletter
\DeclareRobustCommand\rvdots{%
  \vbox{%
    \baselineskip4\p@\lineskiplimit\z@%
    \kern-\p@%
    \hbox{.}\hbox{.}\hbox{.}%
  }%
}

\usepackage{tabularx,environ}
\makeatletter
\newcounter{problem}[section]
\renewcommand{\theproblem}{\arabic{problem}}
\newcommand{\problemtitle}[1]{\gdef\@problemtitle{#1}}%
\newcommand{\probleminput}[1]{\gdef\@probleminput{#1}}%
\newcommand{\problemquestion}[1]{\gdef\@problemquestion{#1}}%
\NewEnviron{problem}{
  \refstepcounter{problem}%
  \problemtitle{}\probleminput{}\problemquestion{}%
  \BODY%
  \par\addvspace{.5\baselineskip}
  \noindent \textbf{Problem \theproblem: \@problemtitle}%
  \par\addvspace{.5\baselineskip}
  \noindent\begin{tabularx}{.5\textwidth}{@{\hspace{\parindent}} l X c}
    \textbf{Input:} & \@probleminput \\
    \textbf{Problem:} & \@problemquestion
  \end{tabularx}
  \par\addvspace{.5\baselineskip}
}
\makeatother
\crefname{problem}{Problem}{Problems}

\newcommand{\ft}[1]{$\mathrm{FT}^{#1}$}

\begin{document}
\title{Optimizing Fault-tolerant Cat State Preparation}

\author{Tom Peham}
\email{tom.peham@tum.de}
\affiliation{Chair for Design Automation, Technical University of Munich, Germany}

\author{Erik Weilandt}
\email{erik.weilandt@tum.de}
\affiliation{Chair for Design Automation, Technical University of Munich, Germany}

\author{Robert Wille}
\email{robert.wille@tum.de}
\affiliation{Chair for Design Automation, Technical University of Munich, Germany}
\affiliation{Munich Quantum Software Company, Germany}

\begin{abstract}
  Cat states are an important resource for fault-tolerant quantum computing, where they serve as building blocks for a variety of fault-tolerant primitives. Consequently, the ability to prepare high-quality cat states at large fault distances is essential.
  While optimizations for low fault distances or small numbers of qubits exist, higher fault distances can be achieved via generalized constructions with potentially suboptimal circuit sizes.
  In this work, we propose a cat state preparation scheme based on preparing two cat states with low-depth circuits, followed by a transversal CNOT and measurement of one of the states.
  This scheme prepares $w$-qubit cat states fault-tolerantly up to fault distances of $9$ using $\ceil{\log_2 w}+1$ depth and at most $3w-2$ CNOTs and $2w$ qubits.
  We discuss that the combinatorially challenging aspect of this construction is the precise wiring of the transversal CNOT and propose three methods for finding these: two based on Satisfiability Modulo Theory solving and one heuristic search based on a local repair strategy.
  Numerical evaluations show that our circuits achieve a high fault-distance while requiring fewer resources as generalized constructions.
\end{abstract}

\maketitle
\section{Introduction}
\label{sec:intro}

Quantum error correction (QEC) and Fault-tolerant quantum computing (FTQC)~\cite{gottesmanTheoryFaulttolerantQuantum1998, shorFaulttolerantQuantumComputation1996, dennisTopologicalQuantumMemory2002, nielsenQuantumComputationQuantum2010} enable a pathway towards utility-scale quantum computing, where billions of operations need to be performed reliably on encoded quantum information~\cite{gidneyHowFactor20482025}.
By redundantly encoding information into quantum error-correcting codes and repeatedly extracting error syndromes, the lifetime of logical qubits can be extended far beyond what is possible for physical qubits.
Recent experimental advances show that current hardware platforms exhibit the capabilities required for FTQC, like real-time, below-threshold decoding~\cite{googlequantumaiandcollaboratorsQuantumErrorCorrection2025}, encoded universal gate sets~\cite{pogorelovExperimentalFaulttolerantCode2024}, and first encoded algorithms~\cite{reichardtLogicalComputationDemonstrated2024}.

Realizing large-scale FTQC requires that every step of a computation be implemented in a fault-tolerant manner, so that physical errors do not spread uncontrollably through the system.
Instead of viewing fault tolerance at the level of an entire algorithm, it is common to decompose computations into subcircuits or \emph{gadgets}~\cite{gottesmanIntroductionQuantumError2009}, each of which is itself fault-tolerant.
If errors are removed quickly via decoding and error correction, they do not accumulate, and information can be protected with high probability.
From this perspective, a central circuit-design task in FTQC is to devise and optimize fault-tolerant gadgets for tasks such as syndrome extraction~\cite{chamberlandFlagFaulttolerantError2018,prabhuFaulttolerantSyndromeExtraction2023, liouReducingQuantumError2024, derksDesigningFaulttolerantCircuits2024}, ancilla state preparation~\cite{zenQuantumCircuitDiscovery2025,pehamAutomatedSynthesisFaultTolerant2025, forlivesiFlagOriginModular2025}, magic-state preparation~\cite{chamberlandVeryLowOverhead2020,gidneyMagicStateCultivation2024,sahayFoldtransversalSurfaceCode2025}, code switching~\cite{buttFaultTolerantCodeSwitchingProtocols2024,heussenEfficientFaulttolerantCode2025}, or \mbox{measurement-free} protocols~\cite{heussenMeasurementFreeFaultTolerantQuantum2024, gotoMeasurementfreeFaulttolerantLogicalzerostate2023}, among others.
While such circuits are often designed manually or for specific code instances, developing general automated compilation and optimization techniques for fault-tolerant circuit constructions has started to gain more and more \mbox{traction~\cite{zenQuantumCircuitDiscovery2025, pehamAutomatedSynthesisFaultTolerant2025, schmidDeterministicFaultTolerantState2025, shuttyDecodingMergedColorSurface2022, rodatzFaultToleranceConstruction2025, rodatzFloquetifyingStabiliserCodes2024}}.

Despite first demonstrations of fault tolerance, state-of-the-art quantum hardware remains strongly resource-constrained~\cite{googlequantumaiandcollaboratorsQuantumErrorCorrection2025, bluvsteinLogicalQuantumProcessor2024, reichardtLogicalComputationDemonstrated2024, postlerDemonstrationFaultTolerantSteane2024, postlerDemonstrationFaulttolerantUniversal2022}.
This means that successful near-term fault-tolerant quantum computing requires careful optimization of fault-tolerant circuits employed during computation.
Even large-scale computations that reuse the same sub-circuits frequently greatly benefit from circuit optimizations, as any optimization to a gadget has an immediate multiplicative impact on resource savings.

This work focuses on optimizing fault-tolerant preparation circuits of specific resource states, so-called \enquote{cat states} which are $w$-qubit states of the form $\frac{1}{\sqrt{2}}(\ket{0}^{\otimes w} + \ket{1}^{\otimes w})$ (also referred to as Greenberg-Horn-Zeilinger states~\cite{greenbergerGoingBellTheorem2007}) or $\frac{1}{\sqrt{2}}(\ket{+}^{\otimes w} + \ket{-}^{\otimes w})$.
Such states are a useful resource for FTQC since they can be used for syndrome extraction~\cite{shorFaulttolerantQuantumComputation1996, tansuwannontAdaptiveSyndromeMeasurements2023,escobar-arrietaImprovedPerformanceBaconShor2025}, magic state preparation~\cite{chamberlandVeryLowOverhead2020}, or fault-tolerant encoders~\cite{sharmaFaultTolerantQuantumLDPC2024}.

Given the conceptually simple and useful nature of these states, it is not surprising that fault-tolerant circuit designs for such states have already been investigated.
Ref.~\cite{prabhuFaulttolerantSyndromeExtraction2023} showed that distance-three cat state preparation can be done with logarithmic overhead, but the overhead increases for preparing cat states with a higher fault-distance.
Ref.~\cite{rodatzFaultToleranceConstruction2025} gives a construction for arbitrary-sized cat states requiring logarithmic depth and linearithmic two-qubit gates and qubits.
Another approach is via flag-fault-tolerant constructions, as any higher-distance flag-fault-tolerant syndrome measurement scheme can be turned into a fault-tolerant cat state preparation scheme requiring worst-case linear depth, two-qubit gate, and qubit overhead~\cite{chamberlandFlagFaulttolerantError2018, forlivesiFlagOriginModular2025}.

\begin{figure*}[!t]
  \centering
  
  \subfloat[]{\includegraphics[width=.23\linewidth]{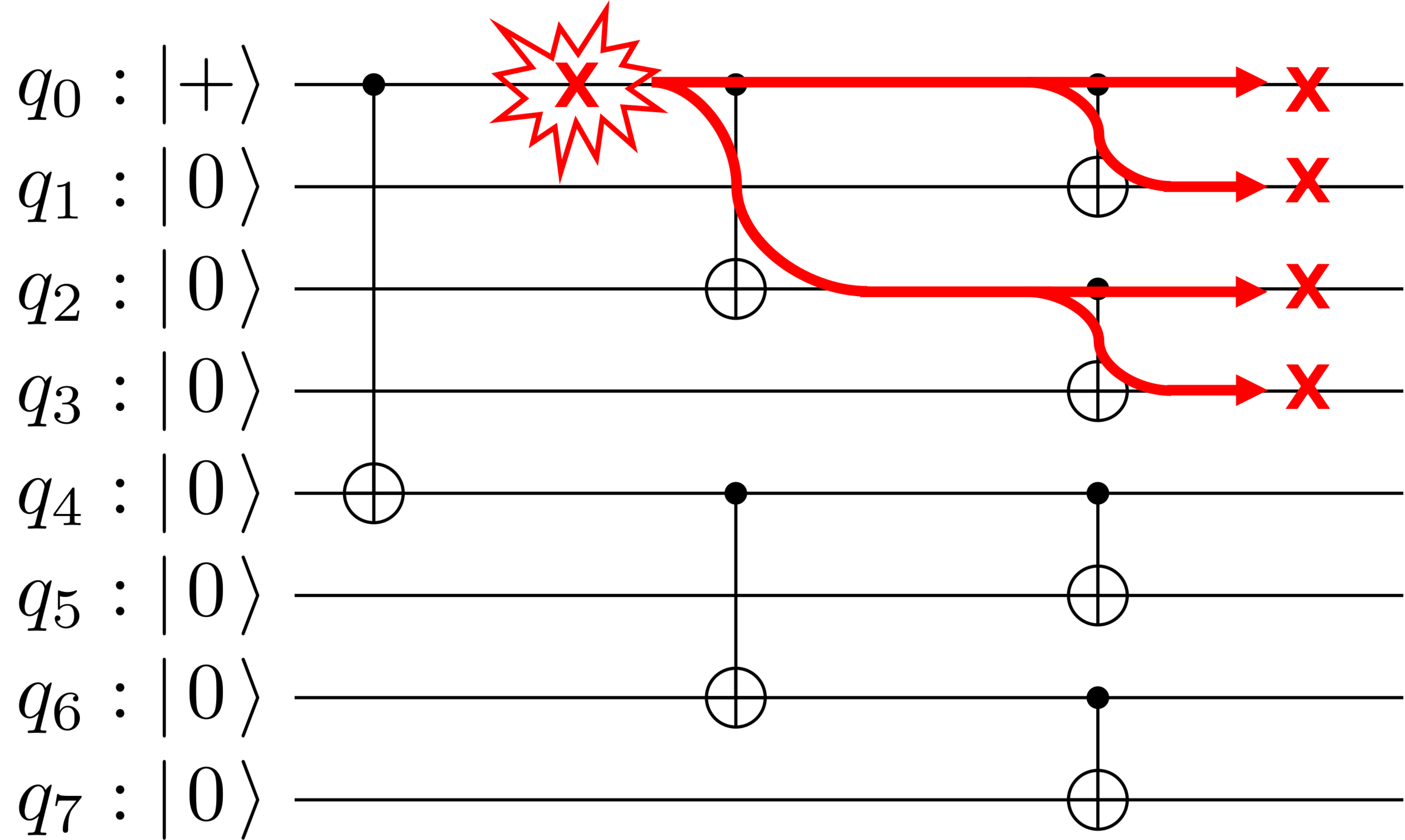}\label{fig:cat:a}}\hfill
  \subfloat[]{\includegraphics[width=.23\linewidth]{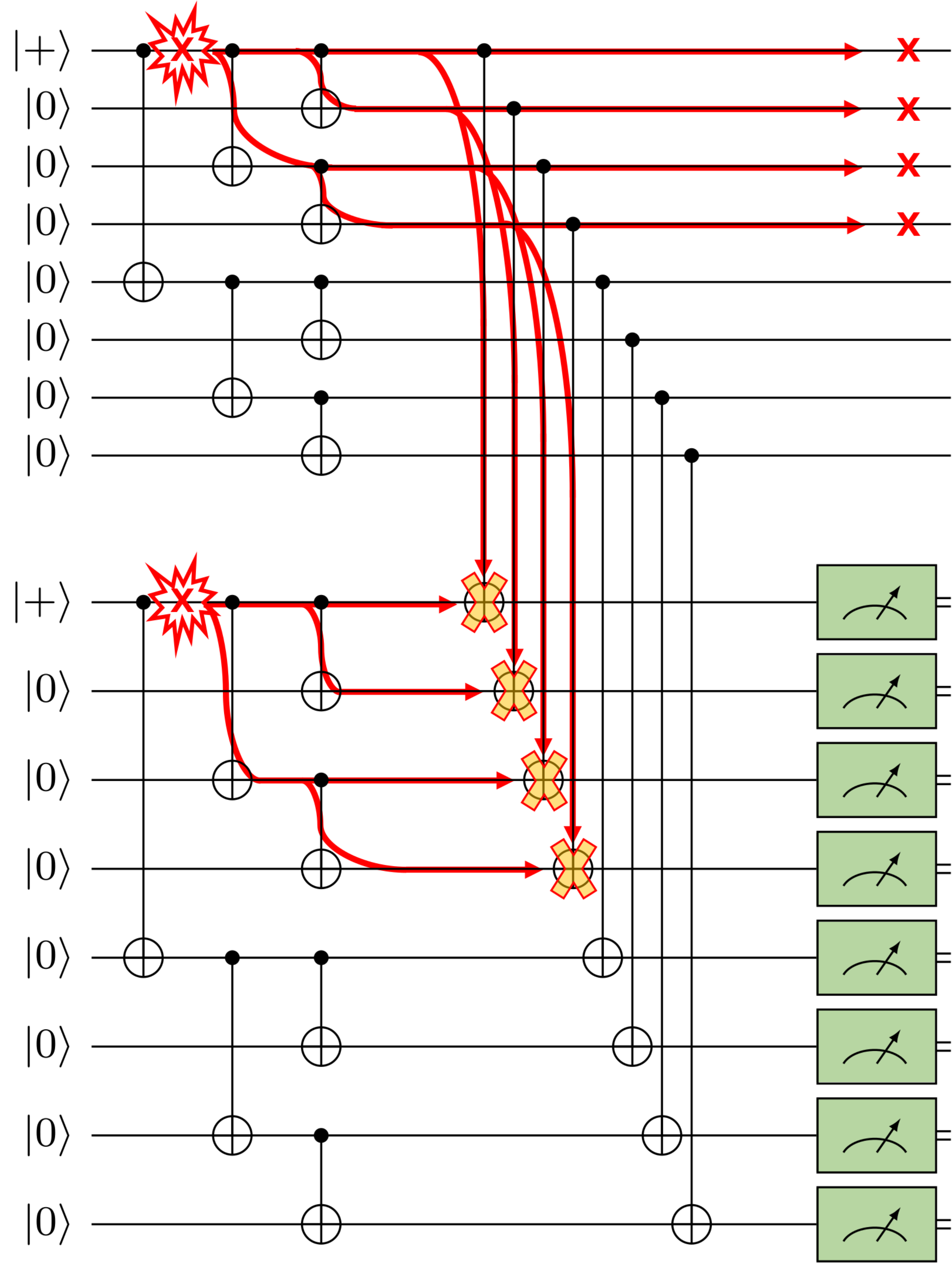}\label{fig:cat:b}}\hfill
  \subfloat[]{\includegraphics[width=.23\linewidth]{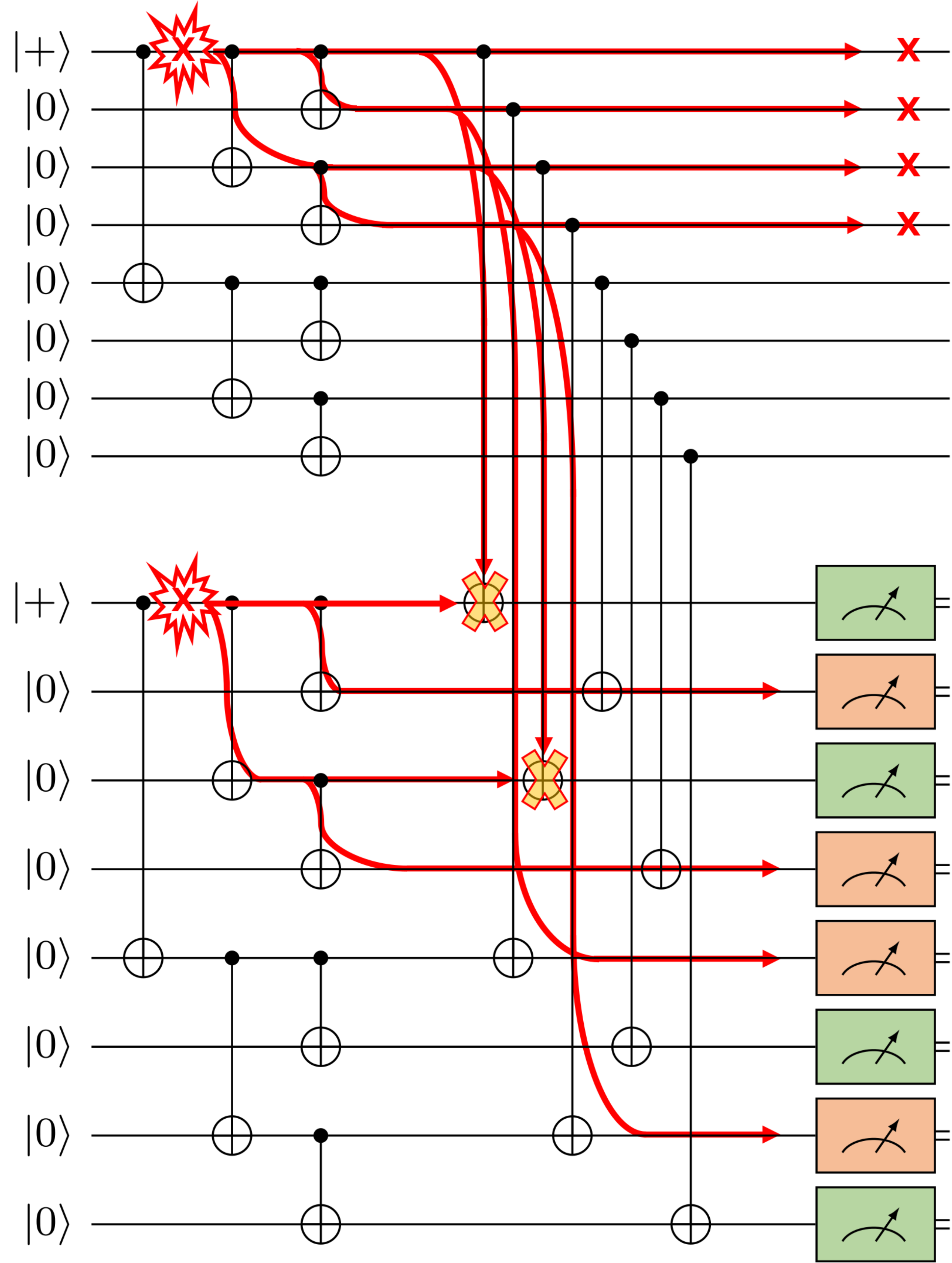}\label{fig:cat:c}}\hfill
  \subfloat[]{\includegraphics[width=.23\linewidth]{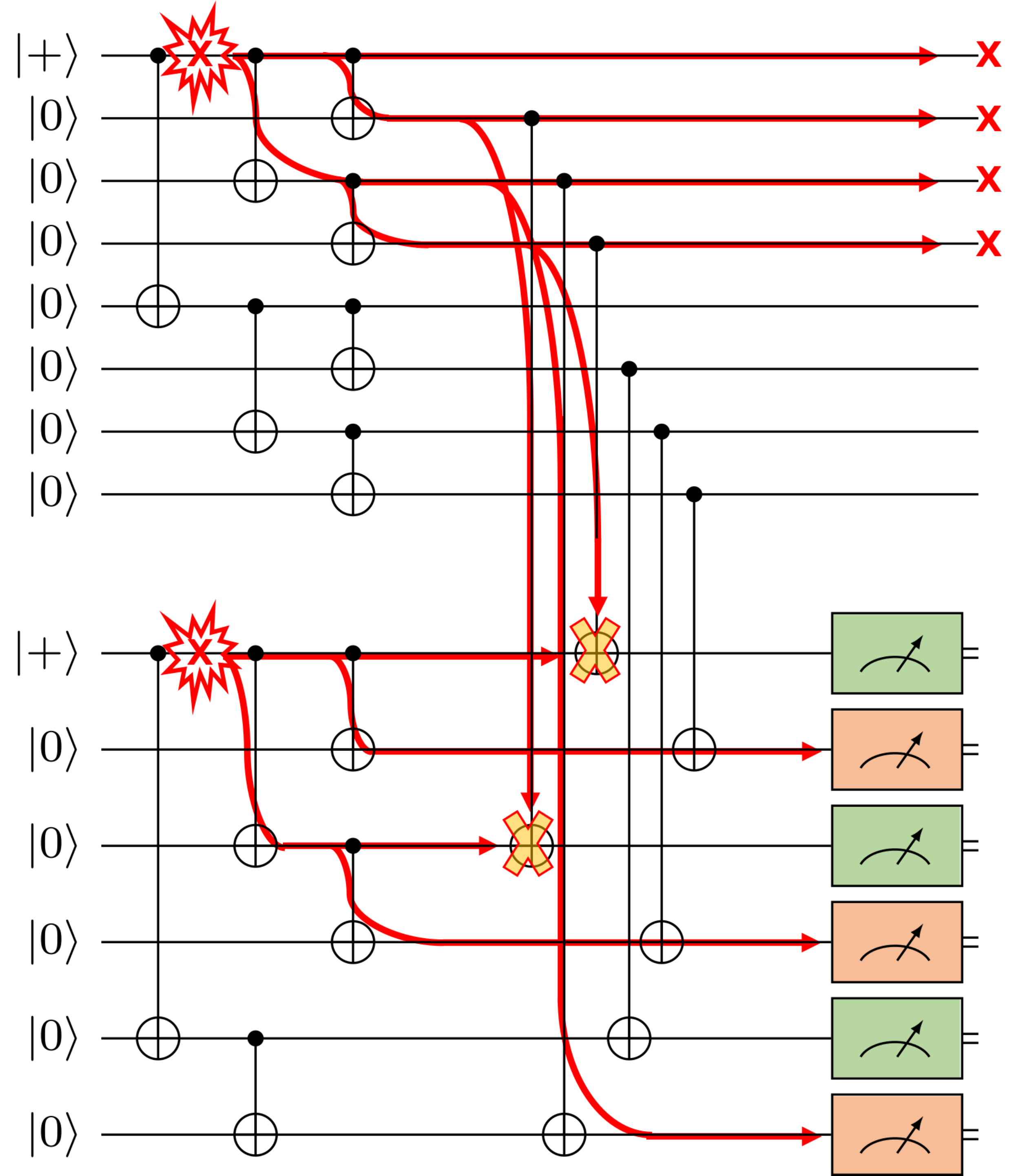}\label{fig:cat:d}}
  
  \caption{Fault-tolerant cat state preparation.
    \textbf{(a)} Non-fault-tolerant log-depth preparation of an 8-qubit cat state. A single weight-1 error propagates to a weight-four error on the prepared state.
    \textbf{(b)} Errors on the data cat state can be detected by an ancilla cat state by copying the $X$ errors to the ancilla and measuring in the $Z$ basis. Here this is not fault-tolerant as two errors are undedected and lead to a weight-four error on the data qubits.
    \textbf{(c)} The error detection gadget can be made fault-tolerant against 4 errors by changing the connectivity of the transversal CNOT.
    \textbf{(d)} Qubit overhead can be reduced by using a smaller ancillary cat state; a 6-qubit cat state with an appropriately connected transversal CNOT is sufficient.}
  \label{fig:cat-state-check}
\end{figure*}

In this work, we aim to further optimize the resource requirements for fault-tolerant cat state preparation.
We propose a scheme for the fault-tolerant preparation of cat states using two (potentially differently-sized) cat states, copying errors from one state to the other, measuring the ancilla state, and post-selecting on the error-free measurement result (see~\Cref{fig:cat-state-check}.
This construction prepares a cat state using logarithmic depth and a linear number of two-qubit gates and qubits.
The challenging part of this construction is finding certain qubit permutations to avoid errors cancelling in such a way as to violate fault tolerance of the preparation schemes.

Our main contributions are as follows:

\begin{itemize}
\item We propose a fault-tolerant cat state preparation scheme that uses two low-depth cat states, a single transversal CNOT layer, and post-selection. 
  This yields $w$-qubit cat states with logarithmic CNOT depth and linear overhead in qubits and two-qubit gates.
  
\item We develop several combinatorial search methods for determining suitable transversal CNOTs that ensure fault tolerance, including a CEGAR-based search, a SAT/SMT formulation, and a heuristic local-repair strategy.
  
\item We apply these methods to synthesize fault-tolerant cat state preparation circuits for up to $49$ qubits for a fault-distance of $4$ and $19$ qubits for a fault distance of $9$.
  
\item We evaluate the constructed circuits using circuit-level noise simulations.
  The resulting circuits exhibit reduced qubit and gate overheads, as well as improved acceptance and error rates, compared to existing general-purpose constructions.
\end{itemize}

The remainder of this work is structured as follows.
\Cref{sec:background} reviews the necessary background on cat states and fault tolerance.
In \Cref{sec:motivation}, we introduce the fault-tolerant cat state preparation scheme studied in this work and formalize the conditions under which a (partial) transversal CNOT preserves fault tolerance.
\Cref{sec:related-work} reviews existing fault-tolerant cat state preparation constructions.
In \Cref{sec:cats}, we analyze the error-propagation structure of balanced-tree cat state preparation circuits and show how this structure can be exploited to restrict the search space for fault-tolerant constructions and efficiently checking fault tolerance of a given candidate CNOT.
\Cref{sec:comb-search} presents the combinatorial search methods used to synthesize fault-tolerant transversal CNOTs.
Finally, \Cref{sec:eval} evaluates the resulting circuits, and \Cref{sec:conclusion} concludes the paper.

\section{Background}
\label{sec:background}
Throughout this work we use $[i] = \{1,2,\dots, i\}\subseteq \Z$.
The \emph{Pauli} matrices are given by

\begin{align*}
  I=
  \begin{bmatrix}
    1&0\\
    0&1          
  \end{bmatrix}\quad
  X=
  \begin{bmatrix}
    0&1\\
    1&0          
  \end{bmatrix}\quad
  Y=
  \begin{bmatrix}
    0&-i\\
    i&0          
  \end{bmatrix}\quad
  Z=
  \begin{bmatrix}
    1&0\\
    0&-1          
  \end{bmatrix},
\end{align*}
and together form the single-qubit Pauli group, which naturally extends to the Pauli group on multiple qubits.
In this work, we consider Pauli operators that are pure $X$- or pure $Z$-type, i.e., $w$-qubit operators of the form $\prod_{i=1}^w\sigma_i$ with $\sigma_i \in \{I, X\}$ or $\sigma_i \in \{I, Z\}$.

We will focus on the preparation of \mbox{$w$-qubit} GHZ states, that is, states of the form $\frac{1}{\sqrt{2}}(\ket{0}^{\otimes w} + \ket{1}^{\otimes w})$. 
Preparation of $\frac{1}{\sqrt{2}}(\ket{+}^{\otimes w} + \ket{-}^{\otimes w})$ works analogously.
GHZ states are stabilizer states on $w$ qubits defined by the stabilizers $X^{\otimes n}$ and $Z_i\otimes Z_{i+1}, i \in [w-1]$.

Any $Z$-type operator with an odd support size anti-commutes with the $X^{\otimes n}$ operator and an arbitrary $X$-type operator anti-commutes with at least one of the $Z$-stabilizers.
Therefore, one can detect up to $w-1$ $X$ (bit flip) errors and any odd number of $Z$ (phase flip) error on the state.
Since any two-qubit $Z$ error is a stabilizer, any $Z$ error is stabilizer equivalent to a trivial error or a single-qubit error.

A \emph{fault} is an error event occurring during the execution of a circuit, and these faults can propagate through the circuit.
$X$-faults propagate through the control of a CNOT gate, i.e., if the control qubit flips, the target qubit does too.
\begin{center}
  \includegraphics[width=0.3\linewidth]{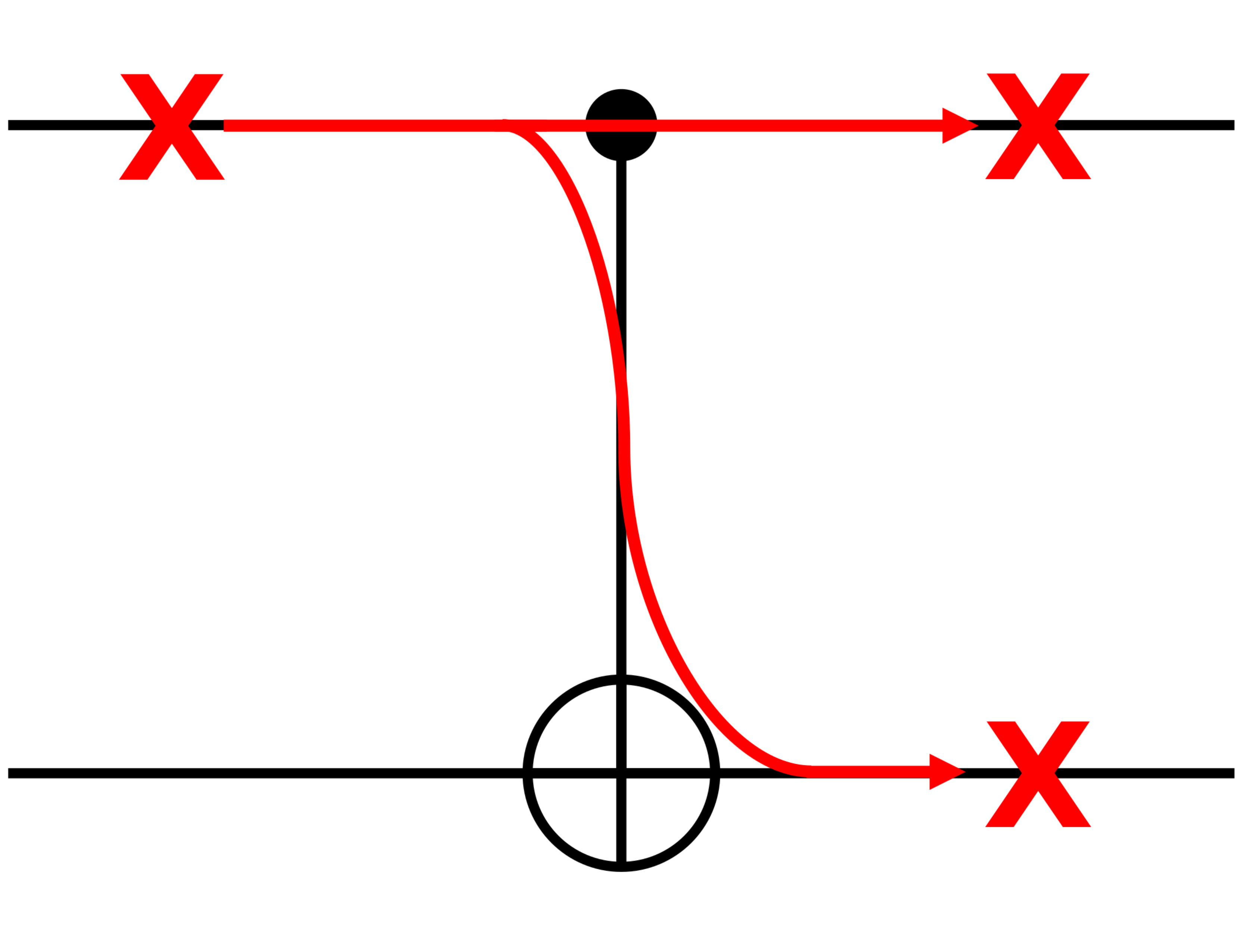}      
\end{center}
An \emph{error} is the fault propagated to the end of the circuit.
We denote an arbitrary $w$-qubit $X$-error as an $\F_2^w$ vector where the vector has non-zero entries exactly on the indices corresponding to the support of the error, e.g., the vector $[1,0,1]$ corresponds to the operator $XIX$.
For a GHZ state, any $X$-error $e \in \F_2^w$ is equivalent to the error $\mathbbm{1}^{\otimes w}+e$ obtained by flipping all bits in $e$ since $X^{\otimes w}$ is a stabilizer of the state.
The \emph{weight} of an error $e$ on a cat state (denoted $\mathrm{wt}~e$) is defined as the number of non-zero entries in $e$ or $\mathbbm{1}+e$, whichever is smaller.

Multiple faults can occur simultaneously in a circuit, causing multiple errors.
If two faults cause errors $e_1$ and $e_2$, the resulting combined error is given by $e_1+e_2$.
The set of errors caused by up to $t$ faults in a circuit $\calC$ is denoted as $\calE_t(\calC)$, which we call the ($t$-) \emph{fault set} of the circuit $\calC$.
If the circuit in question is clear from context, we will often only write $\calE_t$ for the $t$-fault set.

Errors in a circuit can be detected via certain measurements that do not disturb the state.

\begin{definition}
  We say that a cat state preparation scheme is fault-tolerant up to $t$ errors, \ft{t}, if any combination of up to $t$ faults in the circuit causing an error $e \in \calE_t$ on the resulting cat state with weight greater than $t$ is detected.
\end{definition}

\section{Cat State Verification via Transversal CNOTs}
\label{sec:motivation}

Since cat states are CSS states, they can be prepared by initializing qubits in $\ket{+}$ followed by a CNOT circuit~\cite{kissingerPhasefreeZXDiagrams2022,pehamAutomatedSynthesisFaultTolerant2025}.
Specifically, since cat states have only a single $X$-type stabilizer, one qubit is prepared in $\ket{+}$ and the rest in $\ket{0}$. 
The following CNOT circuit must then propagate the initial $X$-stabilizer to the other qubits. 
This can be done in $\ceil{\log_2 w}$ depth with a \enquote{balanced tree} CNOT circuit as shown in~\Cref{fig:cat:a}, where the cat state is doubled in size at every layer by entangling the current state with the same number of $\ket{0}$ qubits.
If $w$ is not a power of two, the last layer only entangles $2^{\ceil{\log_2 w}}-w$ qubits.
We refer to this circuit structure as $\calB_w$.

This preparation is not even \ft{1} in general, as single errors on a control in the circuit propagate to an $X$-error on the entire sub-tree, as shown in the preparation of an $8$-qubit cat state using the circuit $\calB_8$ in~\Cref{fig:cat:a}. 

In general, such high-weight propagated errors can be detected using an ancilla cat state by copying the errors via a transversal CNOT and measuring the ancilla cat state in the $Z$-basis. 
The transversal CNOT acts trivially on the cat state as the $X^{\otimes w}$ stabilizers of the \enquote{control} cat state propagate to a stabilizer of the ancilla cat state, and similarly for the $Z$-type stabilizers.
Assuming the ancilla cat state is error-free, copying an error will flip bits in the cat state according to the support of that error.
In the absence of errors, measuring the ancillary cat state in the $Z$-basis yields either the bit string $0^w$ or $1^w$.
Any flipped bit in the measurement result, therefore, reveals the presence of an error.

However, if the ancilla cat state is not error-free, low-weight faults in the circuits may lead to high-weight errors  going undetected.
In particular, if both cat states are prepared using the same circuit structure, any two faults occurring in the same location in both circuits will cancel out after the transversal CNOT.

\begin{example}
  Consider the circuit in~\Cref{fig:cat:b}.
  It prepares two eight-qubit cat states using the circuit $\calB_8$. 
  A transversal CNOT is used to copy errors from the first eight qubits onto the second set of eight qubits. 
  In the depicted scenario, the fault $X_1X_2X_3X_4$ is copied to the ancilla cat state but remains undetected because the same fault occurs during the preparation of the ancilla cat state.
  Since the resulting residual fault on the prepared cat state has weight $4$, the cat state preparation in~\Cref{fig:cat-state-check} is not \ft{2}.
\end{example}

Error cancellations can be avoided by permuting the wiring of the transversal CNOT.
Since any permutation of the qubits of a cat state leaves the state invariant, any transversal CNOT with $\mathrm{CX}_{i,\sigma(w)}$ for arbitrary permutations $\sigma$ in the symmetric group $\calS_w$ acts trivially on the cat states while copying errors from one state to the other.
This fact can be exploited to render the error detection fault-tolerant. 
Intuitively, as long as it can be ensured that high-weight faults do not cancel out or combine to the stabilizer $X^{\otimes w}$ after permuting the qubits, all problematic errors will be detected.

Detecting errors on a cat state using a transversal CNOT has the benefit of only adding a single layer of CNOTs to the preparation.
If the states are prepared using $\calB_w$, then preparation has CNOT-depth $\ceil{\log_2{w}}+1$ using $3w-2$ CNOTs and post-selects on $w$ qubits.

\begin{example}
  Wiring up the transversal CNOT in~\Cref{fig:cat:b} according to
  \[
    \begin{pmatrix}
      1&2&3&4&5&6&7&8\\
      1&5&3&7&2&6&8&4
    \end{pmatrix},
  \]
  prevents the faults from cancelling out.
  In fact, no combination of up to $t<4$ faults in this construction cancels out such that an error of weight greater than $t$ is undetected.
  Thus, the resulting circuit is \ft{4}.
\end{example}

This CNOT and qubit overhead can be further reduced by using a smaller ancillary system. 
If the ancilla state has $1< w' < w$ qubits, then a transversal CNOT between any $w'$ qubits in the first circuit and the second circuit acts trivially on the system.
We can characterise such a partial transversal CNOT by how it connects qubits of the first and second circuit.
Concretely, it is given by a set of CNOT pairs
\[
\overline{\mathrm{CX}} \subseteq [w]\times [w']
\]
with \(|\overline{\mathrm{CX}}| = w'\) such that no index appears more than once on either side.
Equivalently, \(\overline{\mathrm{CX}}\) defines a bijection between the \(w'\) ancilla qubits and a subset of
\(w'\) data qubits.
Such a CNOT has an associated \emph{mapping} $f: \F_2^w \rightarrow \F_2^{w'}$ of errors on $w$ qubits to errors on $w'$ qubits defined on the $i$-th single-qubit error $\hat{e}_i$ as
\[f(\hat{e}_i) =
  \begin{cases}
    \hat{e}_j \text{ if } (i,j) \in \overline{\mathrm{CX}}\\
    \mathbf{0} \text{ else} 
    
  \end{cases}
\]

Error cancellations must still be avoided by connecting the qubits of the two cat states appropriately.
To define what it means for such a cat state preparation circuit to be \ft{t}, we introduce the following definition.

\begin{definition}
  \label{def:ft-cnot}
  Let $\mathcal{C},\mathcal{C}'$ be two cat state preparation circuits over $w$ qubits and $w' \leq w$ qubits, respectively.
  A (partial) transversal CNOT $\overline{\mathrm{CX}}$ with associated mapping $f: \F_2^w \rightarrow \F_2^{w'}$ is called \emph{$T$-fault-tolerant} (or \ft{T}) for $T>0 \in \mathbb{Z}$ if for all $t,t' \geq 0$ with $t+t' \leq T$ there do \emph{not} exist 
  \[
    e \in \calE_t(\mathcal{C}), \qquad
    e' \in \calE_{t'}(\mathcal{C}')
  \]
  such that the combined error
  \[
    f(e) + e' \;\in\; \{ I^{\otimes w'},\, X^{\otimes w'}\}
  \]
  while 
  \[
    \mathrm{wt}(e) > T.
  \]
\end{definition}

\begin{example}
  Consider again the \ft{4} preparation circuit for an $8$-qubit cat state in~\Cref{fig:cat:c}.
  As it turns out, a $6$-qubit ancilla state is sufficient to make this construction \ft{4} by choosing the partial transversal CNOT

  \begin{equation}
    \{(2,3), (3,6), (4,1), (6,5), (7,4), (8,2)\}    
  \end{equation}

  The resulting circuit is shown in~\Cref{fig:cat:d}.
  $6$ qubits is also the smallest possible ancilla state to make this construction \ft{4}.
\end{example}

The objective of this work is precisely to construct such fault-tolerant cat state preparation schemes using (partial) transversal CNOTs.
Note that the size of the circuit only depends on the size of the ancillary state.
Since we want to minimize resource costs, i.e., depth, CNOT count, and post-selection overhead, we want to find the ancilla with the smallest number of qubits possible.

Before discussing how we tackle this optimization problem, we will briefly review related fault-tolerant cat state preparation constructions.

\section{Related Work}
\label{sec:related-work}

\begin{table}[ht]
  \centering
  \caption{Resource scaling of different \ft{t} w-qubit cat state preparation methods.}
  \label{tab:asymptotics}
  \begin{tabular}{llll}
    \hline
    \textbf{Method} & \textbf{Gate count} & \textbf{Depth} & \textbf{Qubits} \\
    \hline
    This method & $\boldsymbol{}{O(w)}$ & $\boldsymbol{O(\log w)}$ & $\boldsymbol{O(w)}$ \\
    Ref.~\cite{rodatzFaultToleranceConstruction2025}   & $O(w \log w)$ & $\boldsymbol{O(\log w)}$ & $O(w \log w)$ \\
    Refs.~\cite{forlivesiFlagOriginModular2025,chamberlandFlagFaulttolerantError2018}      & $\boldsymbol{O(w)}$ & $O(w)$ & $\boldsymbol{O(w)}$ \\
    \hline
  \end{tabular}
\end{table}

Recently, Ref.~\cite{rodatzFaultToleranceConstruction2025} has derived a $w$-fault-tolerant cat state preparation scheme based on rigorous \emph{distance-preserving} rewrite rules on ZX-diagrams~\cite{coeckePicturingQuantumProcesses2018,vandeweteringZXcalculusWorkingQuantum2020}.
The construction works by recursively constructing $w$-qubit cat states from $w/2$ qubit cat states until the base case of 4-qubits is reached, for which a simple construction using five qubits, five CNOTs, and one measurement is taken.
The combination works by measuring $w/2$ $XX$ parity checks between pairs of qubits of cat states of size $w/2$.
Therefore, every combination step requires $w/2$ extra qubits, $w$ CNOT gates, and post-selects on $w/2$ measurements.
In total this method requires $\frac{w}{2} \log_2{w} + \frac{w}{4}$ qubits, $w \log_2{w} - \frac{3}{4} w$ CNOTs and a depth of $2\log_2{w} + 1$ if $w$ is a power of $2$.
If the desired cat state preparation should be \ft{t} with $t<w$, then the requirements are more relaxed, as at most $t$ parity checks need to be made in the recursive combination.

Another construction of note is based on flag-fault-tolerant measurements~\cite{chamberlandFlagFaulttolerantError2018}.
Flag-based constructions work by detecting errors between pairs of CNOTs interacting with an ancilla qubit that act trivially on the system in the absence of errors. 
This method requires initializing ancillas using a depth $w-1$ circuit where the qubit initialized in $\ket{+}$ acts as the control qubit of all CNOTs.
Recently, Ref.~\cite{forlivesiFlagOriginModular2025} has introduced a method for finding optimal error-detecting flag gadgets.
While the construction has been introduced in the context of state preparation of arbitrary CSS codes, it can also be used for cat state preparation. 
They manage to find circuits for up to $t=5$ faults and up to $29$ qubits.
Fully fault-tolerant cat state initialization is therefore possible for up to $10$-qubit cat states (since we want to protect against up to $t=5$ faults).
In that case, the optimal construction requires $16$ qubits ($10$ for the cat state and $6$ for the flags), $21$ CNOTs ($9$ for the cat state preparation and $12$ for the flag construction), and has linear depth.
While this construction post-selects on fewer measurements for $w=10$ than the proposed transversal construction, the depth-overhead is much higher.
Furthermore, we aim to construct larger cat states, which require fault tolerance beyond $t=5$.
Ref.~\cite{chamberlandFlagFaulttolerantError2018} also gives a general $w$-flag construction for this flag-based linear cat state preparation that requires $2w-1$ qubits with resets ($3w-1$ without resets), $3w-3$ CNOTs, and depth $7w-15$ (for larger $w$).
This means that the general construction (almost) matches the transversal cat state preparation except for the linear depth overhead.

A brief comparison of the resource scaling is shown in~\Cref{tab:asymptotics}.
This shows that asymptotically, preparing cat states using transversal verification scale most favorably in all considered metrics. 
In~\Cref{sec:eval} we will see that this is not only an asymptotic advantage but that the proposed cat state preparation method uses fewer resources than the recursive construction from Ref.~\cite{rodatzFaultToleranceConstruction2025} in almost all cases, even in the very low qubit regime.
However, the suggested improvements hinge on the fact that such $t$-FT permutations can actually be constructed.

Fault-tolerant state preparation using multiple copies of a state and a transversal is a well-known construction~\cite{paetznickFaulttolerantAncillaPreparation2013, gongComputationQuantumReedMuller2024, takagiErrorRatesResource2017}.
In fact our setting is very similar to Ref.~\cite{gongComputationQuantumReedMuller2024} and Ref.~\cite{paetznickFaulttolerantAncillaPreparation2013} where fault-tolerant preparation of logical basis states of the quantum Reed Muller codes and the Golay code are achieved by constructing multiple copies of a state and finding an appropriate permutation of the transversal CNOTs.
The codes considered in those works do not allow all possible permutations of qubits and are more restrictive.
Furthermore, unlike in our case, the ancilla states in those constructions are fixed as no smaller states can be used for verification.
The constructions in the two works are limited to distance $7$ and $9$ codes which translates to a fault distance of $3$ and $4$, respectively.
We aim to construct circuits with very high fault distance and utilize as few qubits as possible.
To achieve this, we need to develop efficient combinatorial search algorithms to construct the desired cat state preparation circuits.
In the next sections, we outline how we achieve this goal.

\section{Constructing Fault-Tolerant Cat State Preparation Circuits}
\label{sec:cats}

Given two cat state preparation circuits $\calC$, $\calC'$ of size $w, w'$ with $w' \leq w$, the task is to construct a partial transversal CNOT that is fault-tolerant in the sense of~\Cref{def:ft-cnot}. %
Since all $w'$ qubits of the ancillary state serve as targets in this construction, the transversal CNOT can be characterised by
\begin{itemize}
\item a choice of $w'$ control qubits on the data cat state, and
\item a permutation of the target qubits on the ancilla cat state.
\end{itemize}

Only certain combinations of these two choices avoid error cancellations that would violate fault tolerance.
Because the number of possible fault combinations grows rapidly with the fault distance, and because the joint search space of size ${w\choose w'}w'!$ is enormous, we rely on structural properties of the preparation circuit to guide the construction.

\subsection{Fault Structure of the Preparation Circuit $\calB_w$}
\label{sec:fault-structure}
A key benefit of using the balanced-tree preparation circuit $\calB_w$ is that its fault structure is highly regular.
Any single $X$ error occurring before a control in $\calB_w$ propagates to the entire binary subtree below it.
For $w=8$, for example, the circuit has the $1$-fault set 
\begin{align*}
  \calE_1 = \{&X_1,X_2,X_3,X_4,X_5,X_6,X_7,X_8, \\
              &X_1X_2, X_3X_4, X_5X_6, X_7X_8, \\
              &X_1X_2X_3X_4, X_5X_6X_7X_8\},
\end{align*}
In general, $\calE_1(\calB_w)$ consists of all $X$-errors supported on dyadic intervals of $[w]$, and thus has size $2w-2$.
The higher-order fault sets $\calE_t(\calB_w)$ are formed by adding these error patterns over $\F_2$.
Due to the regular structure, many such combinations will coincide, thereby avoiding the storage of many redundant elements in memory.
Nevertheless, the size of the $t$-fault set still grows exponentially in $t$.

The fault set structure is essential for two reasons:
\begin{enumerate}
\item it allows us to reason about which errors can propagate to the ancilla and cancel there, and
\item it enables an efficient check of whether a candidate partial transversal CNOT is fault-tolerant.
\end{enumerate}

\subsection{Determining $w'$ and a Choice of Control Qubits}
\label{sec:controls}

In the following, we assume that both $\calC$ and $\calC'$ are implemented using the balanced-tree circuit $\calB_w$ (or $\calB_{w'}$), as introduced in \Cref{sec:motivation}.
Before searching for a permutation of the ancilla targets, we must decide how many ancilla qubits are needed and which data qubits should act as controls.    
This is hard to decide in general, but we can narrow down the search due to the structure of the possible faults.

\begin{figure}[t]
  \centering
  \subfloat[Non-\ft{2} permutation. Two faults lead to a weight four fault on the data.]{\includegraphics[width=.48\linewidth]{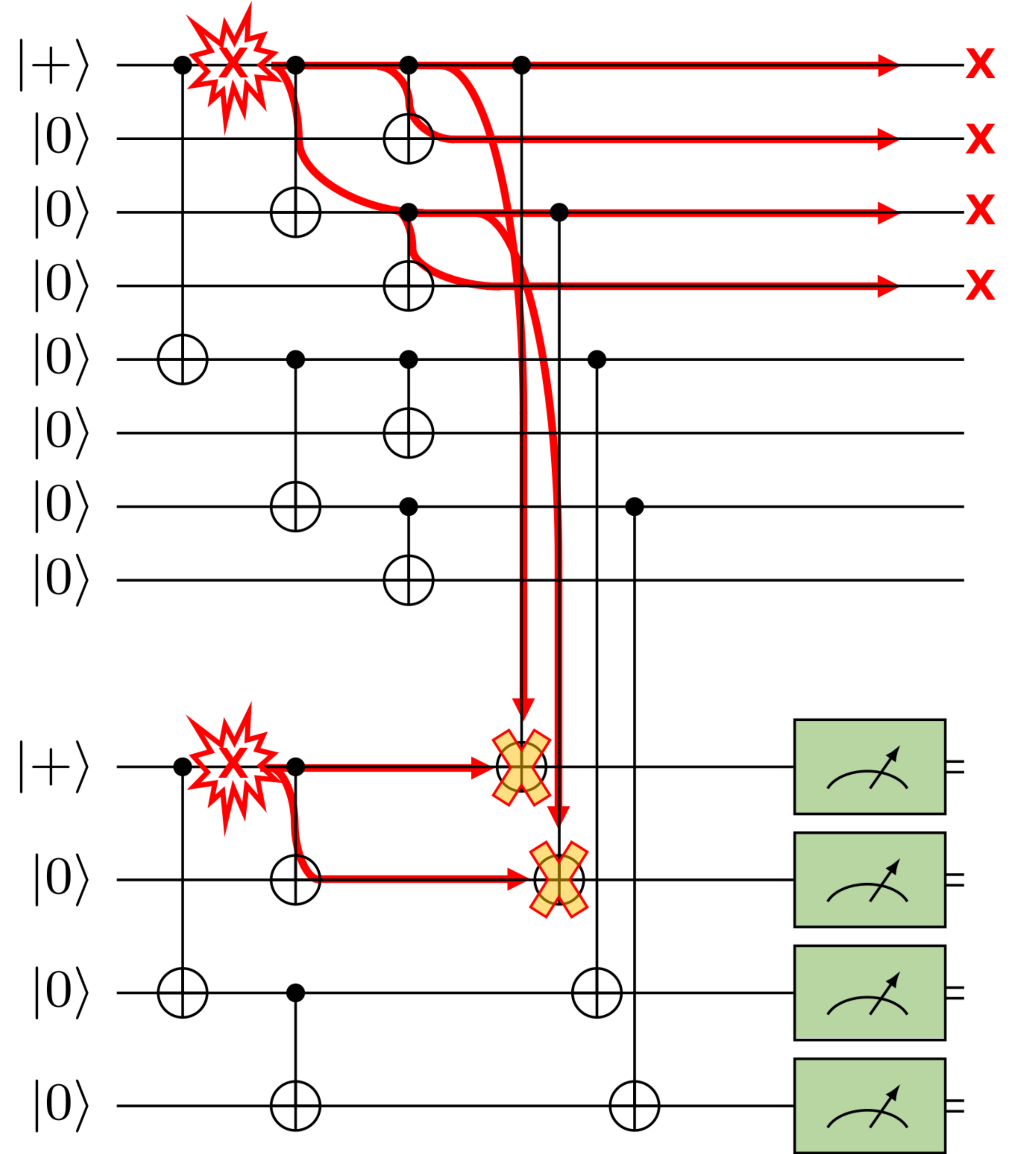}\label{fig:ft2:wrong}}
  \subfloat[\ft{2} permutation.]{\includegraphics[width=.48\linewidth]{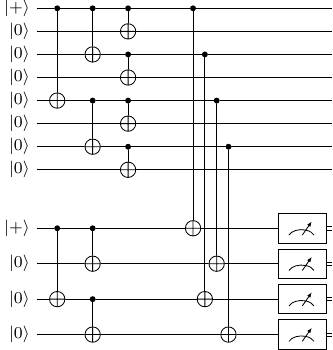}\label{fig:ft2:correct}}
  \caption{\ft{2} cat state preparation.}
  \label{fig:ft-2}
\end{figure}

Let us consider the simplest case first, namely constructing \ft{2} cat state preparation circuits.
If the ancilla had fewer than $w/2$ qubits, then two qubits participating in the same CNOT of the last layer of the data circuit would not be connected to the ancilla, making certain weight-$2$ propagated errors completely undetectable.
Thus $w' \geq w/2$ is necessary, and in fact already sufficient. With $w/2$ controls, each size-$2$ subtree is connected to the ancilla, and the cancellations shown in~\Cref{fig:ft-2} can be avoided by ensuring that entire subtrees never map to entire subtrees.

The size of a subtree can also be used to determine the minimal sizes of the required ancilla states.
Consider, for example, a subtree with four qubits, which means that a single error can lead to a weight-four error on the data qubits.
If only one or two of the qubits are connected to the ancilla state, up to two single-qubit errors are sufficient to cancel the copied errors on the ancilla.
This is completely independent of the target permutation.
Therefore, to achieve \ft{3}, at least three of those data qubits must be connected to the ancilla state.
Since this has to be true for a size-four subtree, we know that \ft{3} cat state preparation requires an ancilla state of size at least $(3/4)w$.

This argument straightforwardly generalizes to larger subtrees as well.
If there is a subtree of size eight, then \ft{7} preparation requires that at least $7$ of those qubits be connected to the ancilla state, and \ft{7} preparation of (large enough) cat states therefore requires at least a $(7/8)w$ qubit ancilla.

Slightly larger ancillas may be needed once propagated ancilla faults are also taken into account, but these lower bounds greatly simplify the search for a possible choice of ancilla size and control qubits.

\subsection{Checking Fault Tolerance for a Fixed Set of Controls}
\label{sec:ft}
Once a candidate ancilla size $w'$ and a set of control positions have been selected, the remaining task is to find a permutation of the $w'$ ancilla qubits that yields an \ft{t} transversal CNOT.
To search efficiently over permutations, we need a fast check for fault tolerance.

Let $\pi: \F_2^w \rightarrow \F_2^{w'}$ be the projection of an error onto the chosen control qubits.
For each $k$, we precompute the projected fault set
\[\calP_k = \pi(\calE_k).\]
Many errors in $\calE_k$ might have the same projection, so for each projected pattern $e'\in \calP_k$, we retain only a maximal-weight representative
\[\calR_k = \{\argmax_{e \in \pi^{-1}(e')}\mathrm{wt}(e) \mid e' \in \calP_k\}.\]
These representatives capture exactly the error patterns that could compromise fault tolerance.
Storing only $\calR_k$ reduces the memory overhead significantly, as many errors in $\calE_k$ are projected identically, and we only keep the worst-case pre-image of each projection.

An error $e \in \calR_k$ violates fault tolerance precisely if
\begin{itemize}
\item fewer than $\wt(e)-k-1$ faults in $\calC'$ are sufficient to cancel $\pi(e)$, and
\item that many faults are still allowed to occur, i.e., $\wt(e)-k-1 \le t-k$.
\end{itemize}
This means that fault tolerance is violated if $\pi(e)$ can be cancelled by at most
\[
  h_k(e) \coloneq \min(t-k,\; \wt(e)-k-1)
\]
faults on the ancilla.

We therefore precompute, for each $e \in \calR_k$, the set of \emph{bad images}:
\[\mathrm{Bad}(e)=\{e' \in \bigcup_{k' = 1}^{h_k(e)} \calE_{k'}(\calC') \mid \wt(e') = \wt(e)\},\]
which are the errors in the ancilla state that could potentially cancel a projected data error under permutation.

Given this set, checking whether a permutation $\sigma$ is \ft{t} reduces to verifying that
\[
  \sigma(\pi(e)) \notin \mathrm{Bad}(e)
  \quad\text{for all } k < t \text{ and all } e \in \calR_k.
\]

As long as we can store the sets $\calR_k$ for each $k$ and $\mathrm{Bad}(e)$ for each error in these sets, checking whether a permutation is fault-tolerant can be done efficiently for each error.

\begin{figure}[t]
  \centering
  \includegraphics[width=.7\linewidth]{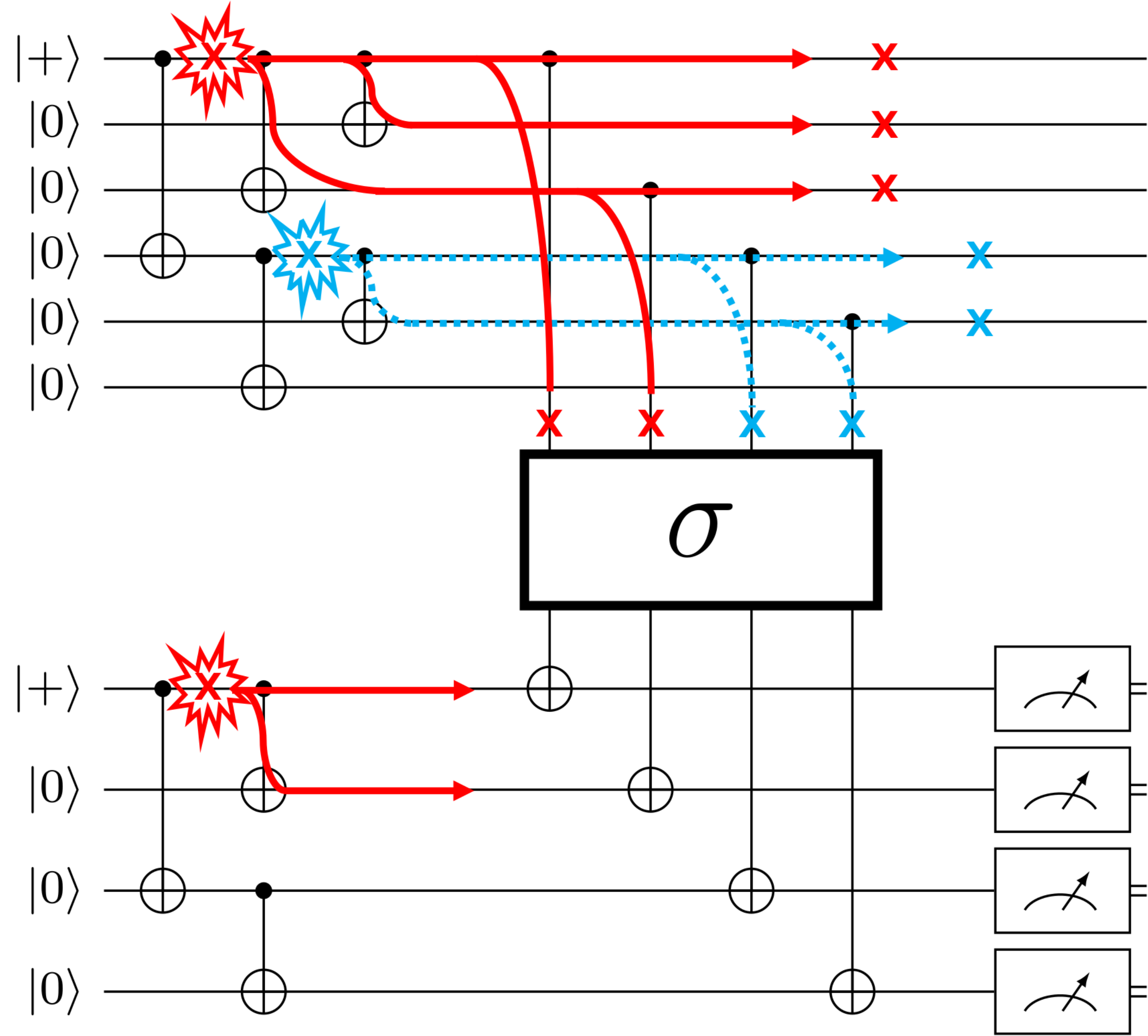}
  \caption{Fault-tolerance check: precomputing what errors on the ancilla can cancel with the worst-case errors on the data qubit allows for fast checking of fault-tolerance based on the permutation $\sigma$.}
  \label{fig:ft-check}
\end{figure}

\begin{example}
  Consider the preparation of a $w=6$ qubit cat state shown in~\Cref{fig:ft-check}.
  Here, the qubits $0$, $2$, $3$, and $4$ have already been selected as control qubits.
  The fault set (up to stabilizer) of that $6$-qubit preparation circuit is
  \begin{align*}
    \{&[0,0,0,0,0,1], [0,0,0,0,1,0], [0,0,0,1,0,0],\\
      &[0,0,1,0,0,0], [0,1,0,0,0,0], [1,0,0,0,0,0], \\
      &[1,1,0,0,0,0], [0,0,0,1,1,0], [1,1,1,0,0,0]\}.
  \end{align*}
  
  Projecting this onto the controls gives the projected fault set
  \begin{align*}
    \{[1,0,0,0], [0,1,0,0], [0,0,1,0], [0,0,0,1], [1,1,0,0]\}.
  \end{align*}
  The only high-weight fault in this projected set is $[1,1,0,0]$, which is the projection of $[0,0,0,1,1,0]$ (drawn in blue in~\Cref{fig:ft-check}) or $[1,1,1,0,0,0]$ (drawn in red in~\Cref{fig:ft-check}).
  This means that in this case $\calR_1$ is simply $\{[1,1,1,0,0,0]\}$.
  The projection of $[1,1,1,0,0,0]$ has weight two, so the bad images are all weight-two faults in the single-qubit fault set of the ancilla preparation circuit:
  \begin{align*}
    \mathrm{Bad}([1,1,1,0,0,0]) = \{[1,1,0,0], [0,0,1,1]\}.
  \end{align*}
  With this, we can quickly check candidates for the permutation $\sigma$ that makes the preparation \ft{2}.
  For example, the permutation
  \[
    \sigma=\begin{pmatrix}
      0&1&2&3\\
      1&0&3&2
    \end{pmatrix}
  \]
  does not work because $\sigma$ projects $[1,1,1,0,0,0]$ onto \mbox{$[1,1,0,0] \in \mathrm{Bad}([1,1,1,0,0,0])$}.
  On the other hand
  \[
    \sigma=\begin{pmatrix}
      0&1&2&3\\
      0&2&1&3
    \end{pmatrix}
  \]
  does work.
  
\end{example}

With a way to narrow down the search for the choice of controls and an efficient method for checking fault-tolerance of a candidate permutation, we now turn to the task of actually constructing \ft{t} partial transversal CNOTs.

\section{Combinatorial Search Methods}
\label{sec:comb-search}

While we are particularly interested in fault-tolerant cat state preparation circuits using a balanced binary tree structure, the methods presented in this section apply to arbitrary cat state preparation circuits $\calC, \calC'$.
The only information required is the single-qubit fault sets $\calE_1(\calC)$ and $\calE_1(\calC')$ of the two cat state preparation circuits, as well as the target fault-distance $t$.

While the circuit structure $\calB_w$ has the lowest possible depth for a unitary cat state preparation circuit, additional (e.g., hardware-) constraints might necessitate other preparation circuits.
In fact, any binary tree on $w$ qubits can be turned into a cat state preparation circuit, and the single-qubit fault set is immediately given by the set of proper subtrees.

Based on these single-qubit fault sets, we can attempt to search for fault-tolerant partial CNOTs.

\subsection{Direct Search for Target Permutation using Satisfiability-Modulo-Theories}
\label{sec:smt}

For a fixed choice of the \(w'\) control positions on the \(w\)-qubit data cat state, the task is to determine how these controls are wired to the \(w'\)-qubit ancilla cat state.
We encode this problem as a satisfiability instance over integer variables and solve it using an SMT solver~\cite{biereHandbookSatisfiability2009,leonardodemouraZ3EfficientSMT2008}.

The wiring is represented by a symbolic permutation
\(\sigma = (\sigma_1,\dots,\sigma_{w'})\),
where each variable \(\sigma_i\) takes values in \([w']\) and specifies the ancilla qubit targeted by the \(i\)-th control.
To ensure that each ancilla qubit is targeted exactly once, the tuple
\((\sigma_1,\dots,\sigma_{w'})\)
is constrained to be a bijection:
\[
  \bigwedge_{i \in [w']} \bigwedge_{j \in [w']\setminus \{i\}} \sigma_i \neq \sigma_j.
\]

We now describe how a projected error pattern on the control qubits propagates through such a permutation.
Any projected error is represented as a bit string \mbox{\(e \in \{0,1\}^{w'}\)}, with support
\(S(e)=\{i \mid e_i=1\}\)
and weight \(|e|\).
Given an assignment to the permutation variables \(\sigma\), the induced error on the ancilla is another bit string
\(e^\sigma \in \{0,1\}^{w'}\),
whose entries are defined by
\[
  e^\sigma_j \;\;\Leftrightarrow\;\;
  \bigvee_{i \in S(e)} \bigl(\sigma_i = j\bigr),
  \qquad j = 1,\dots,w'.
\]
That is, the \(j\)-th ancilla qubit is affected if at least one control qubit in the support of \(e\) is wired to it.
Since \(\sigma\) is a permutation, this mapping preserves Hamming weight:
\[
  \sum_{j=1}^{w'} e^\sigma_j = |e|.
\]

For each error \(e\), let \(\mathrm{Bad}(e) \subseteq \{0,1\}^{w'}\) denote the set of bad images as defined in~\Cref{sec:ft}, each of which has weight \(|e|\).
For each \(e' \in \mathrm{Bad}(e)\), we add the clause
\[
  \bigvee_{j=1}^{w'} \bigl(e'_j \neq e^\sigma_j\bigr),
\]
which forbids the solver from mapping \(e\) to any image in \(\mathrm{Bad}(e)\).

The downside of this encoding is that all forbidden propagations must be encoded upfront, even if a small subset of these propagated faults is sufficient to constrain the choice of permutation. 
Furthermore, the encoding is specific to the choice of control qubits and it might be necessary to solve multiple instances for different choices of control qubits.

\subsection{Simultaneous Search for Control Qubits and Target Permutation with CEGAR}
\label{sec:optimal}

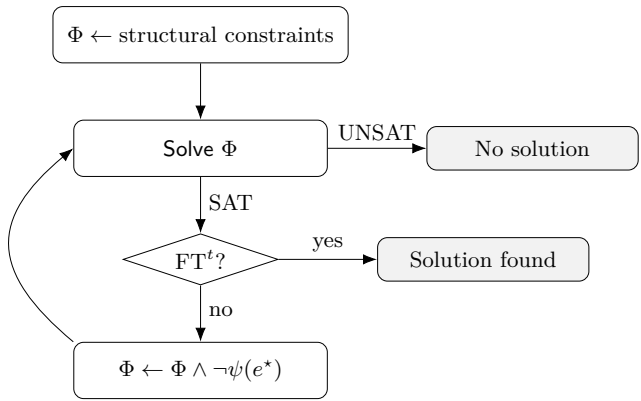
\begin{figure}[t]
  \centering
  \resizebox{\columnwidth}{!}{%
    \begin{tikzpicture}[
      node distance=8mm and 18mm,
      every node/.style={font=\small},
      block/.style    ={draw, rounded corners, align=center, minimum width=36mm, minimum height=8mm, inner sep=2mm},
      decision/.style ={draw, diamond, aspect=2, align=center, inner sep=1pt, minimum width=22mm},
      term/.style     ={draw, rounded corners, align=center, inner sep=2mm, minimum width=30mm, fill=black!5},
      >={Latex[length=2mm]}
      ]
      
      \node[block]     (start)  {$\Phi \gets \text{structural constraints}$};
      \node[block, below=of start] (solve) {$\textsf{Solve } \Phi$};
      \node[decision, below=of solve] (check) {\ft{t}?};
      \node[block, below=of check] (refine) {$\Phi \gets \Phi \land \neg \psi(e^\star)$};
      
      \node[term, right=14mm of solve] (unsat) {No solution};
      \node[term, right=14mm of check] (done)  {Solution found};
      
      \draw[->] (start) -- (solve);
      \draw[->] (solve) -- node[right, pos=.45] {SAT} (check);
      \draw[->] (check) -- node[right, pos=.45] {no} (refine);
      \draw[->] (refine.north west) .. controls +(-12mm,10mm) and +(-12mm,-10mm) .. (solve.west);
      
      \draw[->] (solve) -- node[above, pos=.5] {UNSAT} (unsat);
      \draw[->] (check) -- node[above, pos=.5] {yes} (done);
      
    \end{tikzpicture}%
  }
  \caption{Constructing \ft{t} partial transversal CNOTs for cat state preparation using CEGAR. An initial SAT formula $\Phi$ is iteratively refined by adding blocking clauses forbidding counterexamples until satisfiability is decided.}
  \label{fig:cegar-topdown}
\end{figure}

We now describe a combined exact approach that simultaneously searches over the choice of \(w'\) control qubits among the \(w\) data qubits and the wiring to the \(w'\) ancilla qubits, without encoding all fault propagations upfront.
Because the resulting search space is enormous, we adopt a counterexample-guided abstraction refinement (CEGAR) strategy.

CEGAR~\cite{clarkeCounterexampleguidedAbstractionRefinement2003} is an iterative approach in which one first solves a coarse abstraction of the problem and then checks whether the candidate solution satisfies the full specification.
If a counterexample is found, it is used to refine the abstraction by adding constraints that rule out the offending behavior.
This \emph{solve--check--refine} loop is repeated until either a valid solution is obtained or unsatisfiability is proved.

In our setting, the initial abstraction captures only the structural constraints:
\begin{itemize}
\item exactly \(w'\) data qubits are selected as controls, and
\item the selected controls are wired bijectively to the \(w'\) ancilla qubits.
\end{itemize}

For each data qubit \(q \in [w]\), we introduce a Boolean variable \(\mathrm{ctrl}_q\) indicating whether \(q\) is selected as a control, and an integer variable \(\mathrm{trgt}_q \in \{1,\dots,w'\}\) specifying the ancilla qubit targeted by \(q\) if \(\mathrm{ctrl}_q=1\).
A valid partial transversal CNOT is enforced by the constraints
\[
  \sum_{q=1}^w \mathrm{ctrl}_q = w',
\]
which ensures that exactly \(w'\) controls are chosen, and
\[
  \bigwedge_{u<v}
  \bigl(
    \mathrm{ctrl}_u = 0 \;\lor\;
    \mathrm{ctrl}_v = 0 \;\lor\;
    \mathrm{trgt}_u \neq \mathrm{trgt}_v
  \bigr),
\]
which enforces that no two selected control qubits are connected to the same ancilla target.

Given a candidate assignment \((\mathrm{ctrl},\mathrm{trgt})\), we construct the corresponding partial transversal CNOT \(\overline{\mathrm{CX}}\) and its induced error mapping \(f\), and check whether the resulting construction is \ft{t}.
If no violation is found, the search terminates with an \ft{t} partial transversal CNOT.

If the checker identifies a counterexample, that is, a projected error $e$ whose induced image
$f(e) \in \mathrm{Bad}(e)$\footnote{Since the choice of control qubits is determined jointly with the wiring, the sets $\mathrm{Bad}(e)$ cannot be fully precomputed. In practice, computing and caching them on demand works well, as valid control selections are typically found early.}
under the current assignment $(\mathrm{ctrl},\mathrm{trgt})$, we refine the abstraction by adding a \emph{blocking clause} that excludes this specific assignment.

Concretely, for each ancilla index $j = 1,\dots,w'$, we introduce the Boolean formula
\[
  b_j(e)
  \;\coloneq\;
  \bigvee_{\substack{q \in [w] \\ \mathrm{ctrl}_q = 1,\, e_q = 1}}
  \bigl(\mathrm{trgt}_q = j\bigr),
\]
which evaluates to true if and only if the projected error has suport on ancilla qubit $j$.
To prevent this projection from recurring, we add the clause
\[
  \bigvee_{j=1}^{w'}
  \bigl(f(e)_j \neq b_j(e)\bigr),
\]
thereby excluding the current assignment.

This refinement process accumulates blocking clauses as new counterexamples are discovered, progressively shrinking the search space until either an \ft{t} partial transversal CNOT is found or unsatisfiability is established.

The advantage of this CEGAR formulation is that it avoids encoding all forbidden projections upfront, and instead learns only those constraints that are required to eliminate observed counterexamples.
In some instances, however, a large number of refinements may be needed before convergence.
In such cases, it can be advantageous to fix the choice of control qubits and directly search for a fault-tolerant wiring, as discussed in the previous section.

\subsection{Heuristic Local Search}
\label{sec:heuristic}

\begin{algorithm}[t]
  \caption{Heuristic search for fault-tolerant partial CNOT}
  \label{alg:search}
  \KwIn{Fault set $\calE$ on $w$ qubits, forbidden images $\mathrm{Bad}(e)$ for all $e \in \calE$, maximum iterations $N_{\text{iter}}$, a projection $\pi: \F_2^w \rightarrow \F_2^{w'}$}
  \KwOut{Permutation $\sigma$ making the overall CNOT \ft{t} or \textsc{fail}}
  \BlankLine
  
  $\sigma \gets$ random permutation\;
  
  \For{$k = 1$ \KwTo $N_{\text{iter}}$}{
    Find $e \in \calE$ with $\sigma(\pi(e)) \in \mathrm{Bad}(e)$\;
    \If{no such $e$}{
      \Return{$\sigma$}\;
    }
    $\mathrm{ones} \gets \mathrm{supp}(\pi(e))$\;
    $\mathrm{zeros} \gets [w'] \setminus \mathrm{supp}(\pi(e))$\;
    \While{$\sigma(\pi(e)) \in \mathrm{Bad}(e)$}{
      randomly pick $i \in \mathrm{ones}$\;
      randomly pick $j \in \mathrm{zeros}$\;
      $\sigma \gets \sigma \circ (i\,j)$\;
    }
  }
  \Return{\textsc{fail}}\;
\end{algorithm}

In addition to the exact SMT encoding(s), we also employ a heuristic local search strategy that operates directly on the space of ancilla permutations.
Remember that for a fixed choice of the $w'$ control positions, the task is to find a permutation of the $w'$ ancilla qubits such that any error $e$ is not projected onto an error in $\mathrm{Bad}(e)$.

The procedure begins with a random permutation of the $w'$ ancilla positions.
For a given permutation, we check whether the permutation of any projected error is in the precomputed bad images $\mathrm{Bad}(e)$.
If such an $e$ exists, we attempt to adjust the permutation so that $e$ is no longer a violation.
This can be done by swapping qubits in the support of $e$ with qubits outside the support.
Note that this is not guaranteed to converge, but in practice, it often eliminates the conflicts quickly.
If the search becomes stuck while trying to eliminate the violation, occasional random permutations are performed in the hope of breaking the cycle in the search.

Pseudocode for this algorithm is shown in~\Cref{alg:search}.
This randomized repair process continues until either a valid permutation is found or the maximum number of iterations is reached.
While the method does not guarantee completeness---there may exist valid permutations that it fails to discover---it often succeeds in practice for moderate system sizes.

A key challenge of this approach is that repairing a conflict for one error $e$
does not guarantee that previously repaired conflicts remain satisfied.
In other words, a swap that eliminates $f(e)\in \mathrm{Bad}(e)$ may introduce a new violation for a different error $e'$.
Our procedure is inspired by the resampling algorithm of Moser and Tardos~\cite{moserConstructiveProofGeneral2010}, originally developed as a constructive proof of the Lovász local lemma, which describes the conditions under which a search procedure for such rare combinatorial objects is successful.
Although the formal conditions of the lemma do not hold in our setting, the same randomized repair philosophy often succeeds in practice. Repeated swaps and occasional random perturbations typically shuffle the permutation into a globally valid configuration before the iteration budget is exhausted.

\subsection{Comparison of the Three Search Approaches}
\label{sec:comp}

\begin{figure*}[t]
  \centering
  \subfloat[\# CNOTs]{\includegraphics[width=0.48\linewidth]{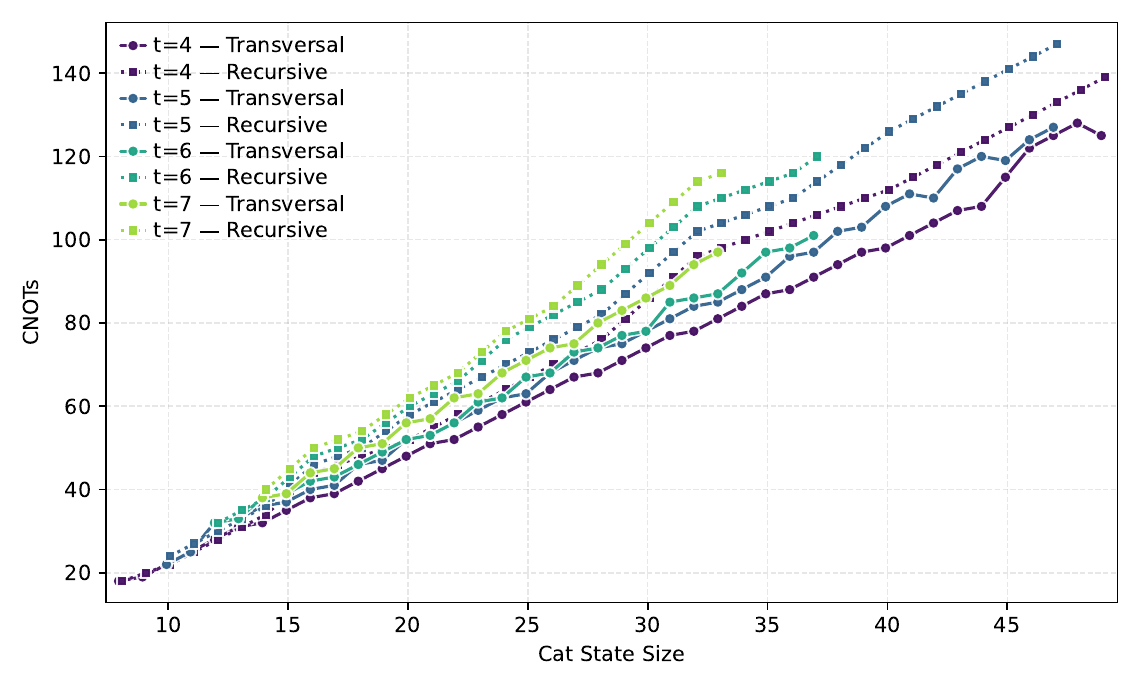}\label{fig:circuit-metrics:cnots}}
  \hfill
  \subfloat[\# Qubits]{\includegraphics[width=0.48\linewidth]{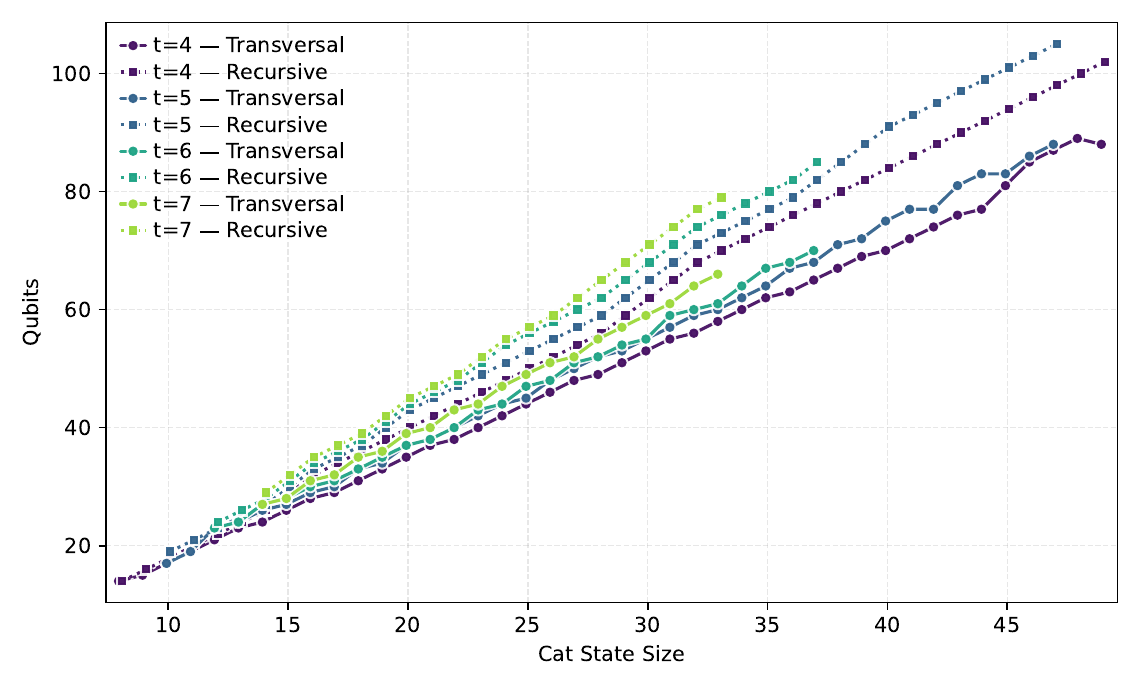}\label{fig:circuit-metrics:qubits}}
  
  \caption{Comparison of Circuit Metrics for Cat State Preparation Circuits for different cat state sizes and fault distance. \enquote{transversal} corresponds to the proposed construction and \enquote{recursive} corresponds to the construction from Ref.~\cite{rodatzFaultToleranceConstruction2025}}
  \label{fig:circuit-metrics}
\end{figure*}

\begin{table}[t]
  \centering
  \caption{Comparison of the three search strategies.}
  \label{tab:method-comparison}
  {\footnotesize
  \setlength{\tabcolsep}{4pt}
  \renewcommand{\arraystretch}{1.1}
  \begin{tabular}{l l}
    \toprule
    \textbf{Method (scope)} & \textbf{Summary} \\
    \midrule
    \begin{minipage}[t]{0.26\columnwidth}
      \raggedright
      \textbf{SAT/SMT}\newline
      \emph{Permutation}
    \end{minipage}
    &
    \begin{minipage}[t]{0.70\columnwidth}
      \raggedright
      \textbf{Pros:} Exact; often proves UNSAT quickly; direct encoding of forbidden errors.\newline
      \textbf{Cons:} Must be repeated for many control sets; typically slowest on SAT instances.
    \end{minipage}
    \\
    \midrule
    \begin{minipage}[t]{0.26\columnwidth}
      \raggedright
      \textbf{CEGAR}\newline
      \emph{Controls + permutation}
    \end{minipage}
    &
    \begin{minipage}[t]{0.70\columnwidth}
      \raggedright
      \textbf{Pros:} Can prove SAT/UNSAT; refinement keeps formulas small.\newline
      \textbf{Cons:} May require many refinements to prove UNSAT.
    \end{minipage}
    \\
    \midrule
    \begin{minipage}[t]{0.26\columnwidth}
      \raggedright
      \textbf{Local search}\newline
      \emph{Permutation}
    \end{minipage}
    &
    \begin{minipage}[t]{0.70\columnwidth}
      \raggedright
      \textbf{Pros:} Fast; often finds a valid permutation quickly in practice.\newline
      \textbf{Cons:} Heuristic; no completeness or UNSAT guarantees.
    \end{minipage}
    \\
    \bottomrule
  \end{tabular}
  }
\end{table}

\Cref{tab:method-comparison} shows an overview of the benefits and downsides of each of the proposed methods.
The CEGAR approach is the most general as it does not presuppose a selection of control qubits for the transversal CNOT.
It can certify both the existence and non-existence of a fault-tolerant partial transversal CNOT for a given $w'$, and can often solve SAT instances much faster than the direct SMT encoding. UNSAT instances, on the other hand, are often slow to solve due to the large number of iterations required.
The SAT/SMT encoding with fixed controls is narrower in scope, but if an instance is UNSAT, this approach can often prove this very quickly in practice.
If the number of different choices of control qubits can be narrowed down, this direct encoding can be very effective for unsatisfiable instances.
The local search heuristic is by far the fastest and often succeeds quickly, but it cannot certify unsatisfiability and may fail to find a fault-tolerant CNOT even if one exists.

In practice, we combine these methods. %
The solver first attempts the joint CEGAR search, and only if this times out do we iterate over control selections and apply the SAT/SMT encoding together with the local search.
Since the direct SMT encoding is fast for UNSAT instances, we use it to quickly filter out choices of controls that do not work, thereby avoiding the need for heuristic search to iterate over permutations unnecessarily if the instance is UNSAT.
This layered strategy balances generality, exactness, and speed: the most powerful method is tried first, while the faster but less reliable methods act as fallbacks when exact approaches run out of time.

\section{Evaluation}
\label{sec:eval}

The proposed search algorithms for \ft{t} transversal CNOTs have been implemented in Python and used to construct fault-tolerant cat state preparation schemes for up to $49$ qubits and fault distances of up to $9$.
In this section, we show the resource costs and circuit-level simulation results of the proposed scheme in comparison with the general recursive cat state preparation construction from~\cite{rodatzFaultToleranceConstruction2025}.
The Python implementation, along with all generated circuits, is available via the open-source MQT QECC repository, which is part of the Munich Quantum Toolkit~\cite{willeMQTHandbookSummary2024}.

\subsection{Simulation Setup}
\label{sec:setup}

\begin{figure*}[t]
  \centering
  \subfloat[Acceptance rates for simulations using a physical error rate of $p=1\%$]{\includegraphics[width=\linewidth]{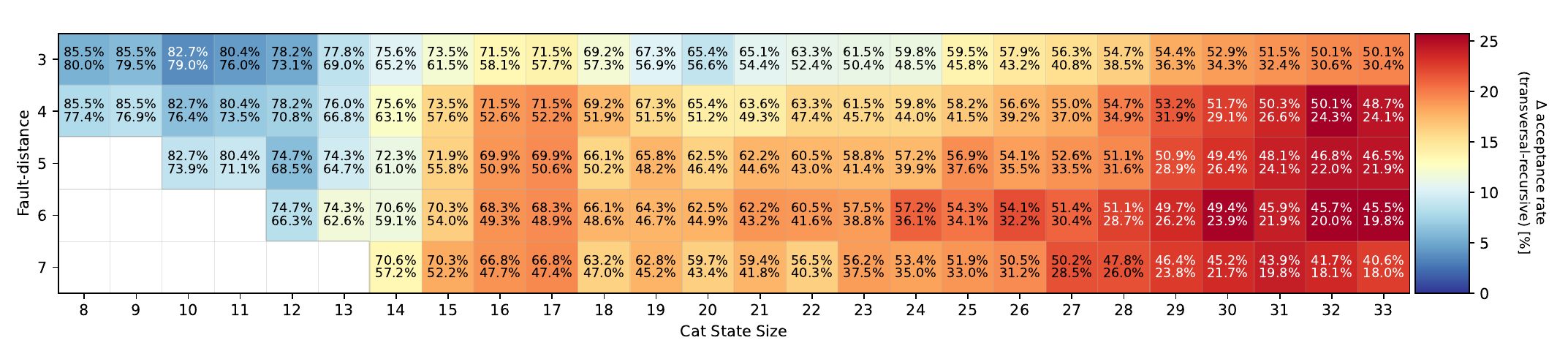}\label{fig:acc-rates:100}}
  
  \subfloat[Acceptance rates for simulations using a physical error rate of $p=0.1\%$]{\includegraphics[width=\linewidth]{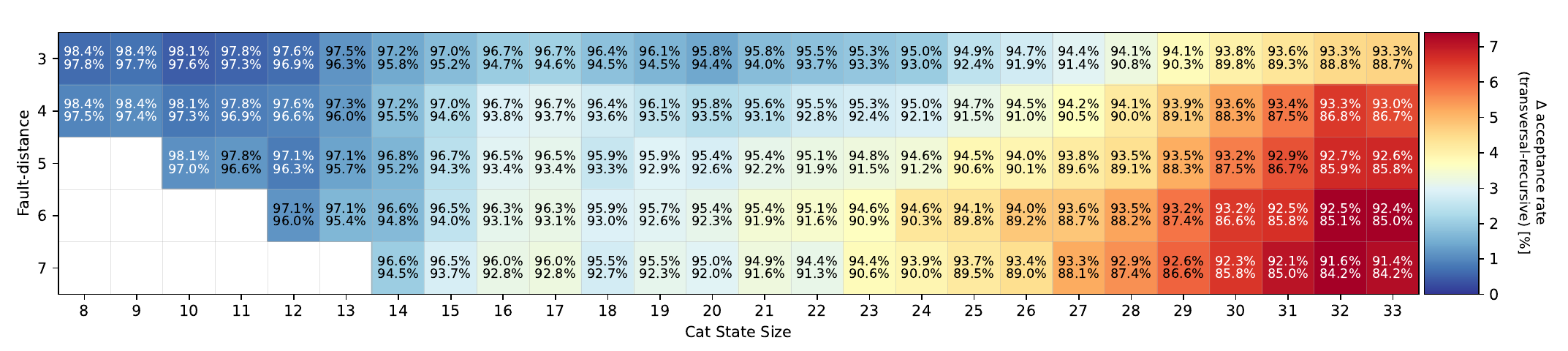}\label{fig:acc-rates:1000}}
  \caption{Comparison of acceptance rates for cat state preparation. \enquote{transversal} corresponds to the proposed construction and \enquote{recursive} corresponds to the construction from Ref.~\cite{rodatzFaultToleranceConstruction2025}.
    Each cell shows the absolute acceptance rate for the transversal (\textbf{top}) and recursive (\textbf{bottom}) construction.
    The hue indicates the difference in acceptance rates of the two constructions.}
  \label{fig:acc-rates}
\end{figure*}

\begin{figure*}[t]
  \centering
  \subfloat[Transversal (this work)]{\includegraphics[width=.48\linewidth]{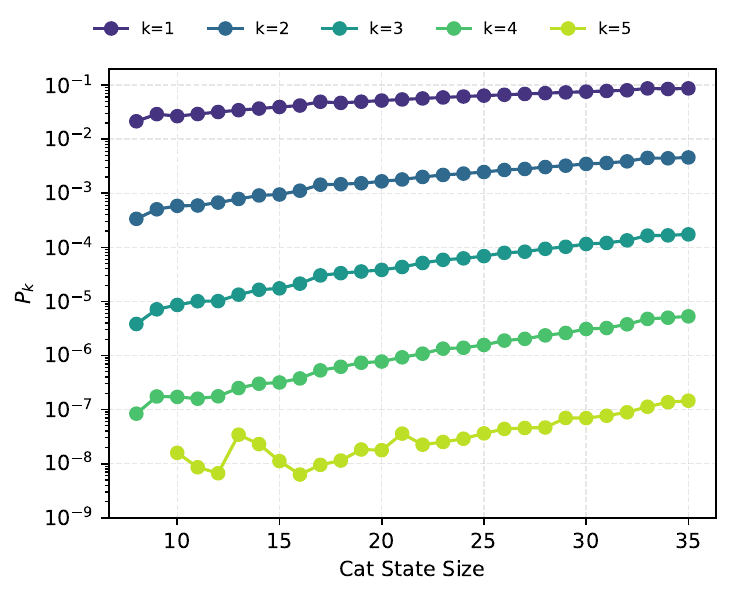}\label{fig:errors_tran_3}}
  \subfloat[Recursive~\cite{rodatzFaultToleranceConstruction2025}]{\includegraphics[width=.48\linewidth]{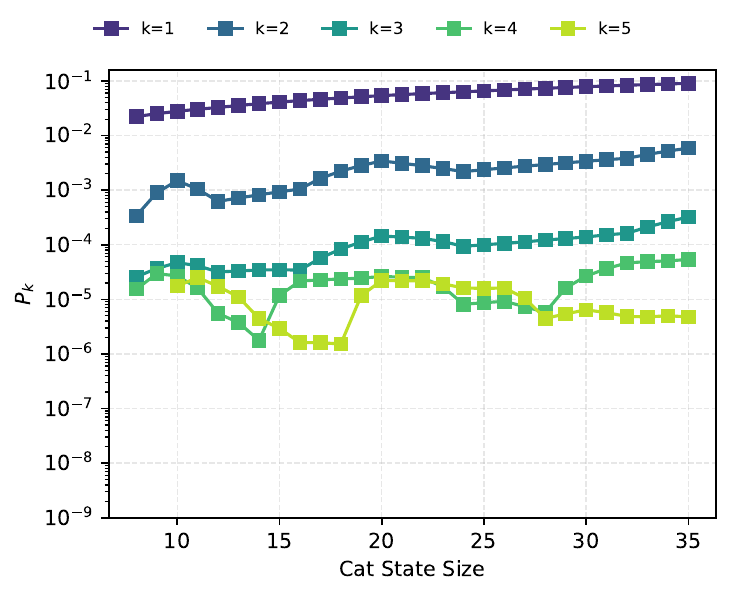}\label{fig:errors_rec_3}}
  \caption{Error rates for \ft{3} cat state preparation at $p=1\%$. $P_k$ is the probability that $k$ bitflips are measured on the data after post-selection.}
  \label{fig:errors_3}
\end{figure*}

The circuits were generated using different machines running Ubuntu 22.04 and required at most \qty{128}{\giga\byte}.

The circuits were simulated using the open-source stabilizer simulator Stim~\cite{gidneyStimFastStabilizer2021} with the following noise model parameterized by a single noise parameter $p$~\footnote{While only bit-flip noise is relevant for our purposes, we employ a standard noise model often used in circuit-level noise simulations to make our simulations more easily comparable to other settings.}:

\begin{itemize}
\item All two-qubit gates are followed by two-qubit depolarizing noise of strength $p$.
\item All initializations flip with a probability $2/3~p$.
\item All measurements flip with a probability $2/3~p$.
\end{itemize}

For the proposed transversal verification scheme, we simulate the entire circuit and post-select on the all-$0$ or all-$1$ $Z$-basis measurement of the ancilla.
If no error is detected on the ancilla, the data is measured out fault-free to obtain a bitstring $b \in \{0,1\}^w$.
Let $n$ be the number of shots simulated, $n_{\mathrm{acc}}$ the number of accepted shots, and $B$ be the set of measured data bitstrings of the accepted shots.
Based on this date, we estimate the acceptance rate $R_\mathrm{acc}$ and the probability $P_k$ that $k$ bits are flipped on the data as

\begin{align*}
  R_\mathrm{acc} &= \frac{n_{\mathrm{acc}}}{n}  \\
  P_k &= \sum_{b \in B} \llbracket (\sum_{i\in w} b_i) = k \rrbracket,
\end{align*}
where $b_i \in \{0,1\}$ is the $i$-th bit of the bit-string $b$ and $\llbracket \cdot \rrbracket$ denotes the \emph{Iverson Bracket} which interprets Bools as integers.

The simulation of the recursive cat state construction proceeds in a similar fashion.

\subsection{Discussion}
\label{sec:discussion}

A comparison of the resulting generated circuits in terms of the number of CNOTs and qubits required to fault-tolerantly prepare a data cat state of a certain size is shown in~\Cref{fig:circuit-metrics}.
We observe that, on both accounts, the proposed transversal construction utilizes fewer resources than the recursive construction.
For larger cat state sizes, up to 20 CNOT gates and qubits can be saved.

This also translates to an improved acceptance rate as can be seen in~\Cref{fig:acc-rates}, which shows the acceptance rates for simulation runs at physical error rates of $1\%$ (\Cref{fig:acc-rates:100}) and $0.1\%$ (\Cref{fig:acc-rates:1000}).
While for high physical error rates of $1\%$ the difference in acceptance rates is as high as $25\%$, for higher-weight cat states, the difference in acceptance rate can still be as high as $6\%$ even for physical error rates of $0.1\%$.
The difference is less significant for lower error rates.

\begin{figure*}[t]
  \centering
  \subfloat[Transversal (this work)]{\includegraphics[width=.48\linewidth]{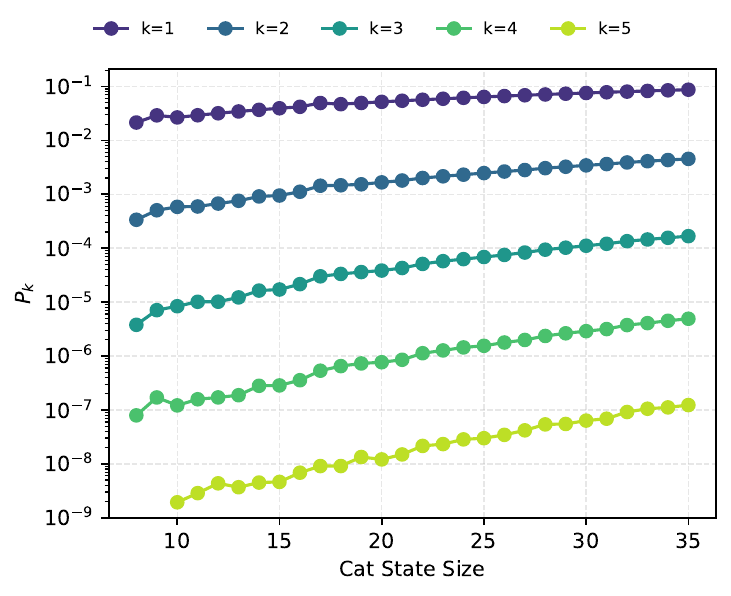}\label{fig:errors_tran_4}}
  \subfloat[Recursive~\cite{rodatzFaultToleranceConstruction2025}]{\includegraphics[width=.48\linewidth]{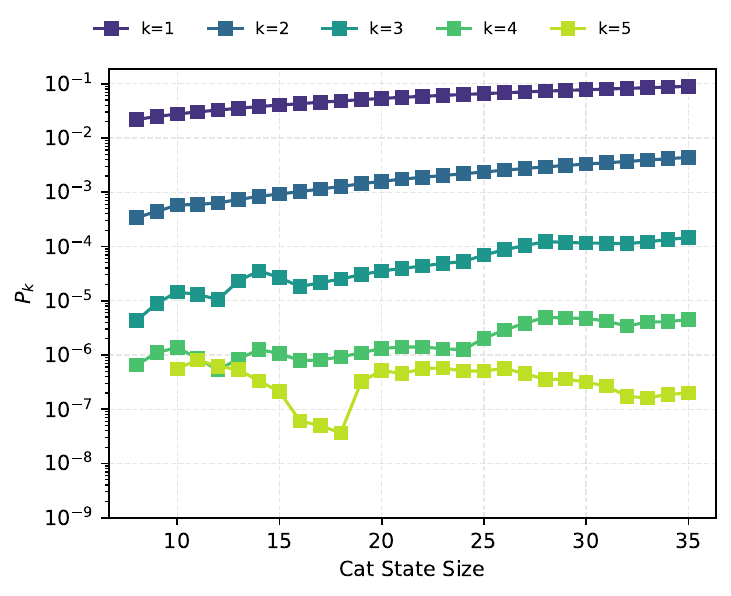}\label{fig:errors_rec_4}}
  \caption{Error rates for \ft{4} cat state preparation at $p=1\%$. $P_k$ is the probability that $k$ bitflips are measured on the data after post-selection.}
  \label{fig:errors_4}
\end{figure*}

In contrast to the acceptance rate, the difference is more pronounced when it comes to errors on the data cat state after post-selection.
Consider the error rate plots in~\Cref{fig:errors_3}, which show simulated error rates for \ft{3} cat states at error rates $p=1\%$.
Since the circuits are constructed to be only fault-tolerant against $3$ errors, we expect the probability that we observe $4$ errors to be about as high as the probability that we observe $3$ errors.
This is also what we see in the simulations of the recursive construction.
However, the transversal verification actually performs much better than that.
We observe that $P_4$ remains approximately an order of magnitude lower than $P_3$. 
Although $P_5$ is not a magnitude lower than $P_4$ for smaller cat states, the scaling is quite favorable compared to the recursive construction.
The same phenomenon can be observed for the \ft{4} (\Cref{fig:errors_4}) cat states, where $5$ bitflips are more likely for the recursive construction compared to the transversal construction.

The reason for this is that while the transversal construction is not guaranteed to catch all problematic combinations of $4$ errors in the \ft{3} case, there are only very few combinations that pass verification undetected.
The recursive construction only checks $3$ qubits of each state when combining them with $ZZ$ measurements and so might miss errors on the other qubits.

Therefore, even if the overhead savings in acceptance rate, depth, qubits, or CNOTs are not substantial for small fault distances, the transversal construction may still provide higher-quality cat states.

More details about the constructed circuits are listed in~\Cref{app:tables}.

\section{Conclusion}
\label{sec:conclusion}

In this work, we propose a method for the fault-tolerant construction of cat states using two cat states, a transversal CNOT gate, and post-selection.
We demonstrate that fault tolerance of this construction depends on the precise connection between the two cat states and the transversal CNOT, and propose three different approaches for finding fault-tolerant CNOTs.
We show that for the parameter ranges for which we are able to solve this combinatorial search problem, our construction outperforms the log-depth construction from Ref.~\cite{rodatzFaultToleranceConstruction2025} in terms of depth, qubit, and CNOT overhead.
Furthermore, circuit-level noise simulations demonstrate that our construction exhibits higher acceptance rates and lower error rates compared to Ref.~\cite{rodatzFaultToleranceConstruction2025}.

While the constructed circuits themselves might have applications in certain Shor-style syndrome extraction schemes~\cite{shorFaulttolerantQuantumComputation1996} or as a resource for other QEC schemes~\cite{escobar-arrietaImprovedPerformanceBaconShor2025}, the proposed search methods might have applications for finding other fault-tolerant constructions.
The CEGAR search in particular might prove useful for other \enquote{brute-force} fault-tolerance constructions as it does not require encoding all constraints up front.

While the approaches of this work and Ref.~\cite{rodatzFaultToleranceConstruction2025} are very different---the former substituting the formal elegance of the latter with raw compute---they complement each other well as the circuits constructed in this work can be used as a starting point for further fault-tolerance construction in fault-tolerant rewriting.
For example, to construct even larger cat states, the optimized circuits of the proposed constructions can be used as part of the recursive construction in Ref.~\cite{rodatzFaultToleranceConstruction2025}.

\section*{Acknowledgements}
The authors would like to thank Lucas Berent and Ludwig Schmid for valuable feedback on the manuscript.

The authors acknowledge funding from the European Research Council (ERC) under the European Union’s Horizon 2020 research and innovation program (grant agreement No.\ 101001318) and Millenion (grant agreement No.\ 101114305). 
This work was part of the Munich Quantum Valley, which is supported by the Bavarian state government with funds from the Hightech Agenda Bayern Plus.
This work was funded by the Deutsche Forschungsgemeinschaft (DFG, German Research Foundation, No. 563402549).

\bibliography{lit_header,zotero}

\clearpage
\appendix

\section{Full List of Circuit Constructions}
\label{app:tables}

\setlength\LTcapwidth{\columnwidth}%
\setlength{\tabcolsep}{3pt}%
\renewcommand{\arraystretch}{1.3}%

\begin{longtable}{r|rrrr|rrrr|rrrr}
\caption{Comparison of \ft{3} cat state preparation protocols at physical error rate $p=0.1\%$.}
\label{tab:t3_p1e-03} \\
\toprule
w & \multicolumn{4}{c}{This Work} & \multicolumn{4}{c}{Recursive~\cite{rodatzFaultToleranceConstruction2025}} & \multicolumn{4}{c}{$\Delta$} \\
 & $R_A$ & d & CX & Q & $R_A$ & d & CX & Q & $R_A$ & d & CX & Q \\
\midrule
8 & 98.44 & 5 & 18 & 14 & 97.78 & 7 & 16 & 13 & 0.66 & -2 & 2 & 1 \\
\rowcolor{black!8} 9 & 98.44 & 6 & 19 & 15 & 97.71 & 7 & 18 & 15 & 0.73 & -1 & 1 & 0 \\
10 & 98.11 & 6 & 22 & 17 & 97.64 & 7 & 20 & 17 & 0.47 & -1 & 2 & 0 \\
\rowcolor{black!8} 11 & 97.84 & 6 & 25 & 19 & 97.27 & 7 & 23 & 19 & 0.57 & -1 & 2 & 0 \\
12 & 97.56 & 6 & 28 & 21 & 96.90 & 7 & 26 & 21 & 0.66 & -1 & 2 & 0 \\
\rowcolor{black!8} 13 & 97.51 & 6 & 29 & 22 & 96.34 & 8 & 29 & 23 & 1.17 & -2 & 0 & -1 \\
14 & 97.24 & 6 & 32 & 24 & 95.78 & 8 & 32 & 25 & 1.46 & -2 & 0 & -1 \\
\rowcolor{black!8} 15 & 96.97 & 6 & 35 & 26 & 95.23 & 8 & 35 & 27 & 1.74 & -2 & 0 & -1 \\
16 & 96.70 & 6 & 38 & 28 & 94.68 & 8 & 38 & 29 & 2.02 & -2 & 0 & -1 \\
\rowcolor{black!8} 17 & 96.70 & 7 & 39 & 29 & 94.61 & 8 & 40 & 31 & 2.09 & -1 & -1 & -2 \\
18 & 96.37 & 7 & 42 & 31 & 94.54 & 8 & 42 & 33 & 1.83 & -1 & 0 & -2 \\
\rowcolor{black!8} 19 & 96.10 & 7 & 45 & 33 & 94.47 & 8 & 44 & 35 & 1.63 & -1 & 1 & -2 \\
20 & 95.84 & 7 & 48 & 35 & 94.40 & 8 & 46 & 37 & 1.43 & -1 & 2 & -2 \\
\rowcolor{black!8} 21 & 95.79 & 7 & 49 & 36 & 94.04 & 8 & 49 & 39 & 1.74 & -1 & 0 & -3 \\
22 & 95.52 & 7 & 52 & 38 & 93.69 & 8 & 52 & 41 & 1.83 & -1 & 0 & -3 \\
\rowcolor{black!8} 23 & 95.25 & 7 & 55 & 40 & 93.33 & 8 & 55 & 43 & 1.92 & -1 & 0 & -3 \\
24 & 94.98 & 7 & 58 & 42 & 92.98 & 8 & 58 & 45 & 2.01 & -1 & 0 & -3 \\
\rowcolor{black!8} 25 & 94.93 & 7 & 59 & 43 & 92.44 & 9 & 61 & 47 & 2.50 & -2 & -2 & -4 \\
26 & 94.67 & 7 & 62 & 45 & 91.91 & 9 & 64 & 49 & 2.76 & -2 & -2 & -4 \\
\rowcolor{black!8} 27 & 94.40 & 7 & 65 & 47 & 91.37 & 9 & 67 & 51 & 3.03 & -2 & -2 & -4 \\
28 & 94.14 & 7 & 68 & 49 & 90.85 & 9 & 70 & 53 & 3.29 & -2 & -2 & -4 \\
\rowcolor{black!8} 29 & 94.09 & 7 & 69 & 50 & 90.32 & 9 & 73 & 55 & 3.77 & -2 & -4 & -5 \\
30 & 93.83 & 7 & 72 & 52 & 89.80 & 9 & 76 & 57 & 4.03 & -2 & -4 & -5 \\
\rowcolor{black!8} 31 & 93.56 & 7 & 75 & 54 & 89.28 & 9 & 79 & 59 & 4.28 & -2 & -4 & -5 \\
32 & 93.30 & 7 & 78 & 56 & 88.76 & 9 & 82 & 61 & 4.54 & -2 & -4 & -5 \\
\rowcolor{black!8} 33 & 93.30 & 8 & 79 & 57 & 88.70 & 9 & 84 & 63 & 4.60 & -1 & -5 & -6 \\
34 & 92.99 & 8 & 82 & 59 & 88.63 & 9 & 86 & 65 & 4.36 & -1 & -4 & -6 \\
\rowcolor{black!8} 35 & 92.73 & 8 & 85 & 61 & 88.57 & 9 & 88 & 67 & 4.16 & -1 & -3 & -6 \\
\bottomrule
\end{longtable}

\pagebreak
\setlength\LTcapwidth{\columnwidth}%
\setlength{\tabcolsep}{3pt}%
\renewcommand{\arraystretch}{1.3}%

\begin{longtable}{r|rrrr|rrrr|rrrr}
\caption{Comparison of \ft{4} cat state preparation protocols at physical error rate $p=0.1\%$.}
\label{tab:t4_p1e-03} \\
\toprule
w & \multicolumn{4}{c}{This Work} & \multicolumn{4}{c}{Recursive~\cite{rodatzFaultToleranceConstruction2025}} & \multicolumn{4}{c}{$\Delta$} \\
 & $R_A$ & d & CX & Q & $R_A$ & d & CX & Q & $R_A$ & d & CX & Q \\
\midrule
8 & 98.44 & 5 & 18 & 14 & 97.46 & 8 & 18 & 14 & 0.98 & -3 & 0 & 0 \\
\rowcolor{black!8} 9 & 98.44 & 6 & 19 & 15 & 97.39 & 8 & 20 & 16 & 1.05 & -2 & -1 & -1 \\
10 & 98.11 & 6 & 22 & 17 & 97.32 & 8 & 22 & 18 & 0.79 & -2 & 0 & -1 \\
\rowcolor{black!8} 11 & 97.84 & 6 & 25 & 19 & 96.95 & 8 & 25 & 20 & 0.89 & -2 & 0 & -1 \\
12 & 97.56 & 6 & 28 & 21 & 96.58 & 8 & 28 & 22 & 0.98 & -2 & 0 & -1 \\
\rowcolor{black!8} 13 & 97.29 & 6 & 31 & 23 & 96.02 & 8 & 31 & 24 & 1.27 & -2 & 0 & -1 \\
14 & 97.24 & 6 & 32 & 24 & 95.47 & 8 & 34 & 26 & 1.77 & -2 & -2 & -2 \\
\rowcolor{black!8} 15 & 96.97 & 6 & 35 & 26 & 94.61 & 9 & 39 & 29 & 2.36 & -3 & -4 & -3 \\
16 & 96.70 & 6 & 38 & 28 & 93.75 & 9 & 44 & 32 & 2.94 & -3 & -6 & -4 \\
\rowcolor{black!8} 17 & 96.70 & 7 & 39 & 29 & 93.68 & 9 & 46 & 34 & 3.01 & -2 & -7 & -5 \\
18 & 96.37 & 7 & 42 & 31 & 93.62 & 9 & 48 & 36 & 2.76 & -2 & -6 & -5 \\
\rowcolor{black!8} 19 & 96.10 & 7 & 45 & 33 & 93.55 & 9 & 50 & 38 & 2.56 & -2 & -5 & -5 \\
20 & 95.84 & 7 & 48 & 35 & 93.48 & 9 & 52 & 40 & 2.36 & -2 & -4 & -5 \\
\rowcolor{black!8} 21 & 95.57 & 7 & 51 & 37 & 93.13 & 9 & 55 & 42 & 2.44 & -2 & -4 & -5 \\
22 & 95.52 & 7 & 52 & 38 & 92.77 & 9 & 58 & 44 & 2.74 & -2 & -6 & -6 \\
\rowcolor{black!8} 23 & 95.25 & 7 & 55 & 40 & 92.42 & 9 & 61 & 46 & 2.83 & -2 & -6 & -6 \\
24 & 94.99 & 7 & 58 & 42 & 92.07 & 9 & 64 & 48 & 2.92 & -2 & -6 & -6 \\
\rowcolor{black!8} 25 & 94.72 & 7 & 61 & 44 & 91.54 & 9 & 67 & 50 & 3.18 & -2 & -6 & -6 \\
26 & 94.45 & 7 & 64 & 46 & 91.01 & 9 & 70 & 52 & 3.44 & -2 & -6 & -6 \\
\rowcolor{black!8} 27 & 94.19 & 7 & 67 & 48 & 90.48 & 9 & 73 & 54 & 3.71 & -2 & -6 & -6 \\
28 & 94.14 & 7 & 68 & 49 & 89.96 & 9 & 76 & 56 & 4.18 & -2 & -8 & -7 \\
\rowcolor{black!8} 29 & 93.88 & 7 & 71 & 51 & 89.15 & 10 & 81 & 59 & 4.73 & -3 & -10 & -8 \\
30 & 93.62 & 7 & 74 & 53 & 88.34 & 10 & 86 & 62 & 5.27 & -3 & -12 & -9 \\
\rowcolor{black!8} 31 & 93.35 & 7 & 77 & 55 & 87.55 & 10 & 91 & 65 & 5.81 & -3 & -14 & -10 \\
32 & 93.30 & 7 & 78 & 56 & 86.76 & 10 & 96 & 68 & 6.54 & -3 & -18 & -12 \\
\rowcolor{black!8} 33 & 93.04 & 8 & 81 & 58 & 86.69 & 10 & 98 & 70 & 6.35 & -2 & -17 & -12 \\
34 & 92.78 & 8 & 84 & 60 & 86.63 & 10 & 100 & 72 & 6.15 & -2 & -16 & -12 \\
\rowcolor{black!8} 35 & 92.52 & 8 & 87 & 62 & 86.57 & 10 & 102 & 74 & 5.96 & -2 & -15 & -12 \\
36 & 92.47 & 8 & 88 & 63 & 86.50 & 10 & 104 & 76 & 5.97 & -2 & -16 & -13 \\
\rowcolor{black!8} 37 & 92.21 & 8 & 91 & 65 & 86.44 & 10 & 106 & 78 & 5.77 & -2 & -15 & -13 \\
38 & 91.96 & 8 & 94 & 67 & 86.38 & 10 & 108 & 80 & 5.58 & -2 & -14 & -13 \\
\rowcolor{black!8} 39 & 91.70 & 8 & 97 & 69 & 86.31 & 10 & 110 & 82 & 5.39 & -2 & -13 & -13 \\
40 & 91.65 & 8 & 98 & 70 & 86.25 & 10 & 112 & 84 & 5.40 & -2 & -14 & -14 \\
\rowcolor{black!8} 41 & 91.40 & 8 & 101 & 72 & 85.92 & 10 & 115 & 86 & 5.47 & -2 & -14 & -14 \\
42 & 91.14 & 8 & 104 & 74 & 85.60 & 10 & 118 & 88 & 5.54 & -2 & -14 & -14 \\
\rowcolor{black!8} 43 & 90.88 & 8 & 107 & 76 & 85.27 & 10 & 121 & 90 & 5.61 & -2 & -14 & -14 \\
44 & 90.84 & 8 & 108 & 77 & 84.95 & 10 & 124 & 92 & 5.89 & -2 & -16 & -15 \\
\rowcolor{black!8} 45 & 90.17 & 8 & 115 & 81 & 84.63 & 10 & 127 & 94 & 5.54 & -2 & -12 & -13 \\
46 & 89.51 & 8 & 122 & 85 & 84.30 & 10 & 130 & 96 & 5.21 & -2 & -8 & -11 \\
\rowcolor{black!8} 47 & 89.26 & 8 & 125 & 87 & 83.98 & 10 & 133 & 98 & 5.28 & -2 & -8 & -11 \\
48 & 89.01 & 8 & 128 & 89 & 83.67 & 10 & 136 & 100 & 5.35 & -2 & -8 & -11 \\
\rowcolor{black!8} 49 & 89.37 & 8 & 125 & 88 & 83.18 & 10 & 139 & 102 & 6.19 & -2 & -14 & -14 \\
\bottomrule
\end{longtable}

\pagebreak
\setlength\LTcapwidth{\columnwidth}%
\setlength{\tabcolsep}{3pt}%
\renewcommand{\arraystretch}{1.3}%

\begin{longtable}{r|rrrr|rrrr|rrrr}
\caption{Comparison of \ft{5} cat state preparation protocols at physical error rate $p=0.1\%$.}
\label{tab:t5_p1e-03} \\
\toprule
w & \multicolumn{4}{c}{This Work} & \multicolumn{4}{c}{Recursive~\cite{rodatzFaultToleranceConstruction2025}} & \multicolumn{4}{c}{$\Delta$} \\
 & $R_A$ & d & CX & Q & $R_A$ & d & CX & Q & $R_A$ & d & CX & Q \\
\midrule
10 & 98.11 & 6 & 22 & 17 & 97.00 & 8 & 24 & 19 & 1.11 & -2 & -2 & -2 \\
\rowcolor{black!8} 11 & 97.84 & 6 & 25 & 19 & 96.63 & 8 & 27 & 21 & 1.20 & -2 & -2 & -2 \\
12 & 97.12 & 6 & 32 & 23 & 96.27 & 8 & 30 & 23 & 0.86 & -2 & 2 & 0 \\
\rowcolor{black!8} 13 & 97.07 & 6 & 33 & 24 & 95.71 & 9 & 33 & 25 & 1.36 & -3 & 0 & -1 \\
14 & 96.80 & 6 & 36 & 26 & 95.16 & 9 & 36 & 27 & 1.64 & -3 & 0 & -1 \\
\rowcolor{black!8} 15 & 96.75 & 6 & 37 & 27 & 94.30 & 9 & 41 & 30 & 2.45 & -3 & -4 & -3 \\
16 & 96.48 & 6 & 40 & 29 & 93.45 & 9 & 46 & 33 & 3.03 & -3 & -6 & -4 \\
\rowcolor{black!8} 17 & 96.48 & 7 & 41 & 30 & 93.38 & 9 & 48 & 35 & 3.10 & -2 & -7 & -5 \\
18 & 95.94 & 7 & 46 & 33 & 93.31 & 9 & 50 & 37 & 2.63 & -2 & -4 & -4 \\
\rowcolor{black!8} 19 & 95.89 & 7 & 47 & 34 & 92.94 & 9 & 54 & 40 & 2.95 & -2 & -7 & -6 \\
20 & 95.40 & 7 & 52 & 37 & 92.57 & 9 & 58 & 43 & 2.84 & -2 & -6 & -6 \\
\rowcolor{black!8} 21 & 95.35 & 7 & 53 & 38 & 92.22 & 9 & 61 & 45 & 3.14 & -2 & -8 & -7 \\
22 & 95.09 & 7 & 56 & 40 & 91.87 & 9 & 64 & 47 & 3.22 & -2 & -8 & -7 \\
\rowcolor{black!8} 23 & 94.82 & 7 & 59 & 42 & 91.52 & 9 & 67 & 49 & 3.30 & -2 & -8 & -7 \\
24 & 94.55 & 7 & 62 & 44 & 91.17 & 9 & 70 & 51 & 3.38 & -2 & -8 & -7 \\
\rowcolor{black!8} 25 & 94.51 & 7 & 63 & 45 & 90.64 & 10 & 73 & 53 & 3.86 & -3 & -10 & -8 \\
26 & 94.03 & 7 & 68 & 48 & 90.12 & 10 & 76 & 55 & 3.90 & -3 & -8 & -7 \\
\rowcolor{black!8} 27 & 93.76 & 7 & 71 & 50 & 89.60 & 10 & 79 & 57 & 4.16 & -3 & -8 & -7 \\
28 & 93.50 & 7 & 74 & 52 & 89.08 & 10 & 82 & 59 & 4.42 & -3 & -8 & -7 \\
\rowcolor{black!8} 29 & 93.45 & 7 & 75 & 53 & 88.28 & 10 & 87 & 62 & 5.17 & -3 & -12 & -9 \\
30 & 93.19 & 7 & 78 & 55 & 87.48 & 10 & 92 & 65 & 5.71 & -3 & -14 & -10 \\
\rowcolor{black!8} 31 & 92.93 & 7 & 81 & 57 & 86.69 & 10 & 97 & 68 & 6.24 & -3 & -16 & -11 \\
32 & 92.67 & 7 & 84 & 59 & 85.91 & 10 & 102 & 71 & 6.76 & -3 & -18 & -12 \\
\rowcolor{black!8} 33 & 92.62 & 8 & 85 & 60 & 85.85 & 10 & 104 & 73 & 6.77 & -2 & -19 & -13 \\
34 & 92.36 & 8 & 88 & 62 & 85.78 & 10 & 106 & 75 & 6.58 & -2 & -18 & -13 \\
\rowcolor{black!8} 35 & 92.10 & 8 & 91 & 64 & 85.72 & 10 & 108 & 77 & 6.38 & -2 & -17 & -13 \\
36 & 91.64 & 8 & 96 & 67 & 85.66 & 10 & 110 & 79 & 5.98 & -2 & -14 & -12 \\
\rowcolor{black!8} 37 & 91.59 & 8 & 97 & 68 & 85.32 & 10 & 114 & 82 & 6.27 & -2 & -17 & -14 \\
38 & 91.13 & 8 & 102 & 71 & 84.98 & 10 & 118 & 85 & 6.15 & -2 & -16 & -14 \\
\rowcolor{black!8} 39 & 91.08 & 8 & 103 & 72 & 84.64 & 10 & 122 & 88 & 6.44 & -2 & -19 & -16 \\
40 & 90.62 & 8 & 108 & 75 & 84.30 & 10 & 126 & 91 & 6.32 & -2 & -18 & -16 \\
\rowcolor{black!8} 41 & 90.36 & 8 & 111 & 77 & 83.98 & 10 & 129 & 93 & 6.38 & -2 & -18 & -16 \\
42 & 90.52 & 8 & 110 & 77 & 83.66 & 10 & 132 & 95 & 6.86 & -2 & -22 & -18 \\
\rowcolor{black!8} 43 & 89.86 & 8 & 117 & 81 & 83.35 & 10 & 135 & 97 & 6.51 & -2 & -18 & -16 \\
44 & 89.61 & 8 & 120 & 83 & 83.03 & 10 & 138 & 99 & 6.58 & -2 & -18 & -16 \\
\rowcolor{black!8} 45 & 89.76 & 8 & 119 & 83 & 82.72 & 10 & 141 & 101 & 7.05 & -2 & -22 & -18 \\
46 & 89.31 & 8 & 124 & 86 & 82.40 & 10 & 144 & 103 & 6.91 & -2 & -20 & -17 \\
\rowcolor{black!8} 47 & 89.06 & 8 & 127 & 88 & 82.09 & 10 & 147 & 105 & 6.97 & -2 & -20 & -17 \\
\bottomrule
\end{longtable}

\pagebreak
\setlength\LTcapwidth{\columnwidth}%
\setlength{\tabcolsep}{3pt}%
\renewcommand{\arraystretch}{1.3}%

\begin{longtable}{r|rrrr|rrrr|rrrr}
\caption{Comparison of \ft{6} cat state preparation protocols at physical error rate $p=0.1\%$.}
\label{tab:t6_p1e-03} \\
\toprule
w & \multicolumn{4}{c}{This Work} & \multicolumn{4}{c}{Recursive~\cite{rodatzFaultToleranceConstruction2025}} & \multicolumn{4}{c}{$\Delta$} \\
 & $R_A$ & d & CX & Q & $R_A$ & d & CX & Q & $R_A$ & d & CX & Q \\
\midrule
12 & 97.12 & 6 & 32 & 23 & 95.95 & 8 & 32 & 24 & 1.17 & -2 & 0 & -1 \\
\rowcolor{black!8} 13 & 97.07 & 6 & 33 & 24 & 95.40 & 9 & 35 & 26 & 1.67 & -3 & -2 & -2 \\
14 & 96.58 & 6 & 38 & 27 & 94.85 & 9 & 38 & 28 & 1.73 & -3 & 0 & -1 \\
\rowcolor{black!8} 15 & 96.53 & 6 & 39 & 28 & 93.99 & 9 & 43 & 31 & 2.54 & -3 & -4 & -3 \\
16 & 96.26 & 6 & 42 & 30 & 93.14 & 9 & 48 & 34 & 3.11 & -3 & -6 & -4 \\
\rowcolor{black!8} 17 & 96.26 & 7 & 43 & 31 & 93.07 & 9 & 50 & 36 & 3.18 & -2 & -7 & -5 \\
18 & 95.94 & 7 & 46 & 33 & 93.01 & 9 & 52 & 38 & 2.93 & -2 & -6 & -5 \\
\rowcolor{black!8} 19 & 95.67 & 7 & 49 & 35 & 92.63 & 9 & 56 & 41 & 3.04 & -2 & -7 & -6 \\
20 & 95.40 & 7 & 52 & 37 & 92.27 & 9 & 60 & 44 & 3.14 & -2 & -8 & -7 \\
\rowcolor{black!8} 21 & 95.35 & 7 & 53 & 38 & 91.92 & 9 & 63 & 46 & 3.44 & -2 & -10 & -8 \\
22 & 95.09 & 7 & 56 & 40 & 91.57 & 9 & 66 & 48 & 3.52 & -2 & -10 & -8 \\
\rowcolor{black!8} 23 & 94.60 & 7 & 61 & 43 & 90.92 & 9 & 71 & 51 & 3.68 & -2 & -10 & -8 \\
24 & 94.55 & 7 & 62 & 44 & 90.28 & 9 & 76 & 54 & 4.27 & -2 & -14 & -10 \\
\rowcolor{black!8} 25 & 94.08 & 7 & 67 & 47 & 89.76 & 10 & 79 & 56 & 4.31 & -3 & -12 & -9 \\
26 & 94.03 & 7 & 68 & 48 & 89.24 & 10 & 82 & 58 & 4.78 & -3 & -14 & -10 \\
\rowcolor{black!8} 27 & 93.55 & 7 & 73 & 51 & 88.73 & 10 & 85 & 60 & 4.83 & -3 & -12 & -9 \\
28 & 93.50 & 7 & 74 & 52 & 88.21 & 10 & 88 & 62 & 5.29 & -3 & -14 & -10 \\
\rowcolor{black!8} 29 & 93.24 & 7 & 77 & 54 & 87.42 & 10 & 93 & 65 & 5.82 & -3 & -16 & -11 \\
30 & 93.19 & 7 & 78 & 55 & 86.63 & 10 & 98 & 68 & 6.56 & -3 & -20 & -13 \\
\rowcolor{black!8} 31 & 92.51 & 7 & 85 & 59 & 85.85 & 10 & 103 & 71 & 6.66 & -3 & -18 & -12 \\
32 & 92.46 & 7 & 86 & 60 & 85.07 & 10 & 108 & 74 & 7.38 & -3 & -22 & -14 \\
\rowcolor{black!8} 33 & 92.41 & 8 & 87 & 61 & 85.01 & 10 & 110 & 76 & 7.40 & -2 & -23 & -15 \\
34 & 91.95 & 8 & 92 & 64 & 84.95 & 10 & 112 & 78 & 7.00 & -2 & -20 & -14 \\
\rowcolor{black!8} 35 & 91.48 & 8 & 97 & 67 & 84.89 & 10 & 114 & 80 & 6.59 & -2 & -17 & -13 \\
36 & 91.43 & 8 & 98 & 68 & 84.82 & 10 & 116 & 82 & 6.61 & -2 & -18 & -14 \\
\rowcolor{black!8} 37 & 91.18 & 8 & 101 & 70 & 84.48 & 10 & 120 & 85 & 6.69 & -2 & -19 & -15 \\
\bottomrule
\end{longtable}

\pagebreak
\noindent
\begin{minipage}{\columnwidth}%
  \vspace*{-\baselineskip} %
  \setlength\LTcapwidth{\columnwidth}%
\setlength{\tabcolsep}{3pt}%
\renewcommand{\arraystretch}{1.3}%

\begin{longtable}{r|rrrr|rrrr|rrrr}
\caption{Comparison of \ft{7} cat state preparation protocols at physical error rate $p=0.1\%$.}
\label{tab:t7_p1e-03} \\
\toprule
w & \multicolumn{4}{c}{This Work} & \multicolumn{4}{c}{Recursive~\cite{rodatzFaultToleranceConstruction2025}} & \multicolumn{4}{c}{$\Delta$} \\
 & $R_A$ & d & CX & Q & $R_A$ & d & CX & Q & $R_A$ & d & CX & Q \\
\midrule
14 & 96.58 & 6 & 38 & 27 & 94.54 & 9 & 40 & 29 & 2.04 & -3 & -2 & -2 \\
\rowcolor{black!8} 15 & 96.53 & 6 & 39 & 28 & 93.68 & 9 & 45 & 32 & 2.84 & -3 & -6 & -4 \\
16 & 96.04 & 6 & 44 & 31 & 92.84 & 9 & 50 & 35 & 3.20 & -3 & -6 & -4 \\
\rowcolor{black!8} 17 & 96.04 & 7 & 45 & 32 & 92.77 & 9 & 52 & 37 & 3.27 & -2 & -7 & -5 \\
18 & 95.50 & 7 & 50 & 35 & 92.70 & 9 & 54 & 39 & 2.80 & -2 & -4 & -4 \\
\rowcolor{black!8} 19 & 95.45 & 7 & 51 & 36 & 92.33 & 9 & 58 & 42 & 3.12 & -2 & -7 & -6 \\
20 & 94.97 & 7 & 56 & 39 & 91.97 & 9 & 62 & 45 & 3.00 & -2 & -6 & -6 \\
\rowcolor{black!8} 21 & 94.92 & 7 & 57 & 40 & 91.62 & 9 & 65 & 47 & 3.30 & -2 & -8 & -7 \\
22 & 94.44 & 7 & 62 & 43 & 91.27 & 9 & 68 & 49 & 3.17 & -2 & -6 & -6 \\
\rowcolor{black!8} 23 & 94.39 & 7 & 63 & 44 & 90.63 & 9 & 73 & 52 & 3.77 & -2 & -10 & -8 \\
24 & 93.91 & 7 & 68 & 47 & 89.99 & 9 & 78 & 55 & 3.93 & -2 & -10 & -8 \\
\rowcolor{black!8} 25 & 93.65 & 7 & 71 & 49 & 89.47 & 10 & 81 & 57 & 4.18 & -3 & -10 & -8 \\
26 & 93.39 & 7 & 74 & 51 & 88.95 & 10 & 84 & 59 & 4.44 & -3 & -10 & -8 \\
\rowcolor{black!8} 27 & 93.34 & 7 & 75 & 52 & 88.15 & 10 & 89 & 62 & 5.19 & -3 & -14 & -10 \\
28 & 92.87 & 7 & 80 & 55 & 87.35 & 10 & 94 & 65 & 5.51 & -3 & -14 & -10 \\
\rowcolor{black!8} 29 & 92.61 & 7 & 83 & 57 & 86.57 & 10 & 99 & 68 & 6.04 & -3 & -16 & -11 \\
30 & 92.35 & 7 & 86 & 59 & 85.78 & 10 & 104 & 71 & 6.57 & -3 & -18 & -12 \\
\rowcolor{black!8} 31 & 92.09 & 7 & 89 & 61 & 85.01 & 10 & 109 & 74 & 7.08 & -3 & -20 & -13 \\
32 & 91.63 & 7 & 94 & 64 & 84.24 & 10 & 114 & 77 & 7.38 & -3 & -20 & -13 \\
\rowcolor{black!8} 33 & 91.37 & 8 & 97 & 66 & 84.18 & 10 & 116 & 79 & 7.19 & -2 & -19 & -13 \\
\bottomrule
\end{longtable}

\end{minipage}
\setlength\LTcapwidth{\columnwidth}%
\setlength{\tabcolsep}{3pt}%
\renewcommand{\arraystretch}{1.3}%

\begin{longtable}{r|rrrr|rrrr|rrrr}
\caption{Comparison of \ft{8} cat state preparation protocols at physical error rate $p=0.1\%$.}
\label{tab:t8_p1e-03} \\
\toprule
w & \multicolumn{4}{c}{This Work} & \multicolumn{4}{c}{Recursive~\cite{rodatzFaultToleranceConstruction2025}} & \multicolumn{4}{c}{$\Delta$} \\
 & $R_A$ & d & CX & Q & $R_A$ & d & CX & Q & $R_A$ & d & CX & Q \\
\midrule
16 & 96.04 & 6 & 44 & 31 & 92.54 & 10 & 52 & 36 & 3.50 & -4 & -8 & -5 \\
\rowcolor{black!8} 17 & 96.04 & 7 & 45 & 32 & 92.47 & 9 & 54 & 38 & 3.57 & -2 & -9 & -6 \\
18 & 95.50 & 7 & 50 & 35 & 92.40 & 9 & 56 & 40 & 3.10 & -2 & -6 & -5 \\
\rowcolor{black!8} 19 & 95.24 & 7 & 53 & 37 & 92.03 & 9 & 60 & 43 & 3.21 & -2 & -7 & -6 \\
20 & 94.76 & 7 & 58 & 40 & 91.66 & 9 & 64 & 46 & 3.09 & -2 & -6 & -6 \\
\bottomrule
\end{longtable}

\setlength\LTcapwidth{\columnwidth}%
\setlength{\tabcolsep}{3pt}%
\renewcommand{\arraystretch}{1.3}%

\begin{longtable}{r|rrrr|rrrr|rrrr}
\caption{Comparison of \ft{9} cat state preparation protocols at physical error rate $p=0.1\%$.}
\label{tab:t9_p1e-03} \\
\toprule
w & \multicolumn{4}{c}{This Work} & \multicolumn{4}{c}{Recursive~\cite{rodatzFaultToleranceConstruction2025}} & \multicolumn{4}{c}{$\Delta$} \\
 & $R_A$ & d & CX & Q & $R_A$ & d & CX & Q & $R_A$ & d & CX & Q \\
\midrule
18 & 95.50 & 7 & 50 & 35 & 92.10 & 10 & 58 & 41 & 3.40 & -3 & -8 & -6 \\
\rowcolor{black!8} 19 & 95.24 & 7 & 53 & 37 & 91.73 & 10 & 62 & 44 & 3.51 & -3 & -9 & -7 \\
\bottomrule
\end{longtable}

\end{document}